\documentclass[acmsmall,screen,nonacm]{acmart}

\newcommand{\smidge}{{\kern .05em}}

\newcommand{\xhdr}[1]{\vspace{1.7mm}\noindent{{\bf #1.}}}
\newcommand{\xhdrNoPeriod}[1]{\vspace{1.7mm}\noindent{{\bf #1}}}

\newcommand{\bnm}{\begin{newmath}}
\newcommand{\enm}{\end{newmath}}
\newcommand{\bne}{\begin{newequation}}
\newcommand{\ene}{\end{newequation}}

\newenvironment{newmath}{\begin{displaymath}%
\setlength{\abovedisplayskip}{4pt}%
\setlength{\belowdisplayskip}{4pt}%
\setlength{\abovedisplayshortskip}{6pt}%
\setlength{\belowdisplayshortskip}{6pt} }{\end{displaymath}}

\newenvironment{newequation}{\begin{equation}%
\setlength{\abovedisplayskip}{4pt}%
\setlength{\belowdisplayskip}{4pt}%
\setlength{\abovedisplayshortskip}{6pt}%
\setlength{\belowdisplayshortskip}{6pt} }{\end{equation}}

\newcommand{\tabref}[1]{Table~\ref{#1}}

\newcommand{\verylongrightarrow}[1]             
      {\setlength{\unitlength}{.01in}           
      \begin{picture}(#1,1) \put(0,0){\vector(1,0){#1}} \end{picture}}

\newlength{\saveparindent}
\newlength{\saveparskip}
\newcounter{ctr}

\definecolor{clr2_1}{RGB}{179,102,158}
\definecolor{clr2_2}{RGB}{152,152,77}

\usepackage{booktabs,subcaption,amsfonts,dcolumn}
\usepackage{diagbox}
\usepackage{hyphenat}
\usepackage{marginnote}

\definecolor{red}{RGB}{246, 114, 127}
\definecolor{blue}{RGB}{108, 86, 123}
\definecolor{orange}{RGB}{150,72,39}

\newcommand{\wordformat}[1]{\textbf{\textit{\textcolor{blue}{#1}}}}
\newcommand{\wordformattable}[1]{\textbf{\tiny{\textcolor{blue}{#1}}}}
\newcommand{\domainformat}[1]{\textbf{\textit{\textcolor{orange}{#1}}}}
\newcommand{\domainformattable}[1]{\textbf{\tiny{\textcolor{orange}{#1}}}}

\begin{document}

\title{Characterizing Alternative Monetization Strategies on YouTube}

\author{Yiqing Hua}
\authornote{Both authors contributed equally to this research.}
\affiliation{
  \institution{Cornell Tech}
  \country{USA}}
\email{yiqing@cs.cornell.edu}

\author{Manoel Horta Ribeiro}
\authornotemark[1]
\affiliation{
  \institution{EPFL}
  \country{Switzerland}}
\email{manoel.hortaribeiro@epfl.ch}

\author{Thomas Ristenpart}
\affiliation{
  \institution{Cornell Tech}
  \country{USA}}
\email{ristenpart@cornell.edu}

\author{Robert West}
\affiliation{
  \institution{EPFL}
  \country{Switzerland}}
\email{robert.west@epfl.ch}

\author{Mor Naaman}
\affiliation{
  \institution{Cornell Tech}
  \country{USA}}
\email{mor.naaman@cornell.edu}

\renewcommand{\shortauthors}{Yiqing Hua et al.}

\begin{abstract}
One of the key emerging roles of the YouTube platform is providing creators the ability to generate revenue from their content and interactions.
Alongside tools provided directly by the platform, such as revenue-sharing from advertising, creators co-opt the platform to use a variety of off-platform monetization opportunities.
In this work, we focus on studying and characterizing these alternative monetization strategies.
Leveraging a large longitudinal YouTube dataset of popular creators, we develop a taxonomy of alternative monetization strategies and a simple methodology to detect their usage automatically. 
We then proceed to characterize the adoption of these strategies.
First, we find that the use of external monetization is expansive and increasingly prevalent, used in 18\% of all videos, with 61\% of channels using one such strategy at least once.
Second, we show that the adoption of these strategies varies substantially among channels of different kinds and popularity, and that channels that establish these alternative revenue streams often become more productive on the platform.
Lastly, we investigate how potentially problematic channels -- those that produce Alt-lite, Alt-right, and Manosphere content -- leverage alternative monetization strategies, finding that they employ a more diverse set of such strategies significantly more often than a carefully chosen comparison set of channels.
This finding complicates YouTube’s role as a gatekeeper, since the practice of excluding policy-violating content from its native on-platform monetization may not be effective.
Overall, this work provides an important step toward broadening the understanding of the monetary incentives behind content creation on YouTube. 

\end{abstract}



\settopmatter{printacmref=false}

\maketitle

\section{Introduction}

YouTube is one of the most popular, important, and consequential technology platforms that exist today. 
According to Pew~\cite{pew1}, in the U.S., around three-quarters of adults use YouTube for diverse reasons that include passing time, deciding whether to buy a particular product, and understanding what is happening in the world---and the U.S. only accounts for less than 20\% of the site's traffic~\cite{alexa}. 
Despite YouTube's influence on the global information ecosystem, the platform downplays its own significant agency by claiming that it is an ``open-armed, egalitarian facilitation of expression, not an elitist gatekeeper with normative and technical restrictions~\cite{gillespie2010politics}.'' 

One of the major roles YouTube plays goes beyond providing a neutral distribution infrastructure.
Since 2008, the platform has shared ad revenues with some content creators through the ``YouTube partner program~\cite{ytmonet}.''
More recently, it added new monetization features such as super-chats (allowing small one-time donations to creators during live streams)~\cite{superchat}.
This business model requires YouTube to navigate between the needs of content creators, end users, and advertisers, while aiming to turn a significant profit~\cite{caplan_tiered_2020}.  
As a result, YouTube has placed guidelines and restrictions about \emph{who} and \emph{what} is allowed to monetize directly on their platform~\cite{ytmonetguidelines}.
Content that violates YouTube's policies is ``demonetized,'' i.e., banned from receiving ad revenue.

\begin{figure}
\centering
\begin{subfigure}[b]{0.48\textwidth}
\centering
\includegraphics[width=\linewidth]{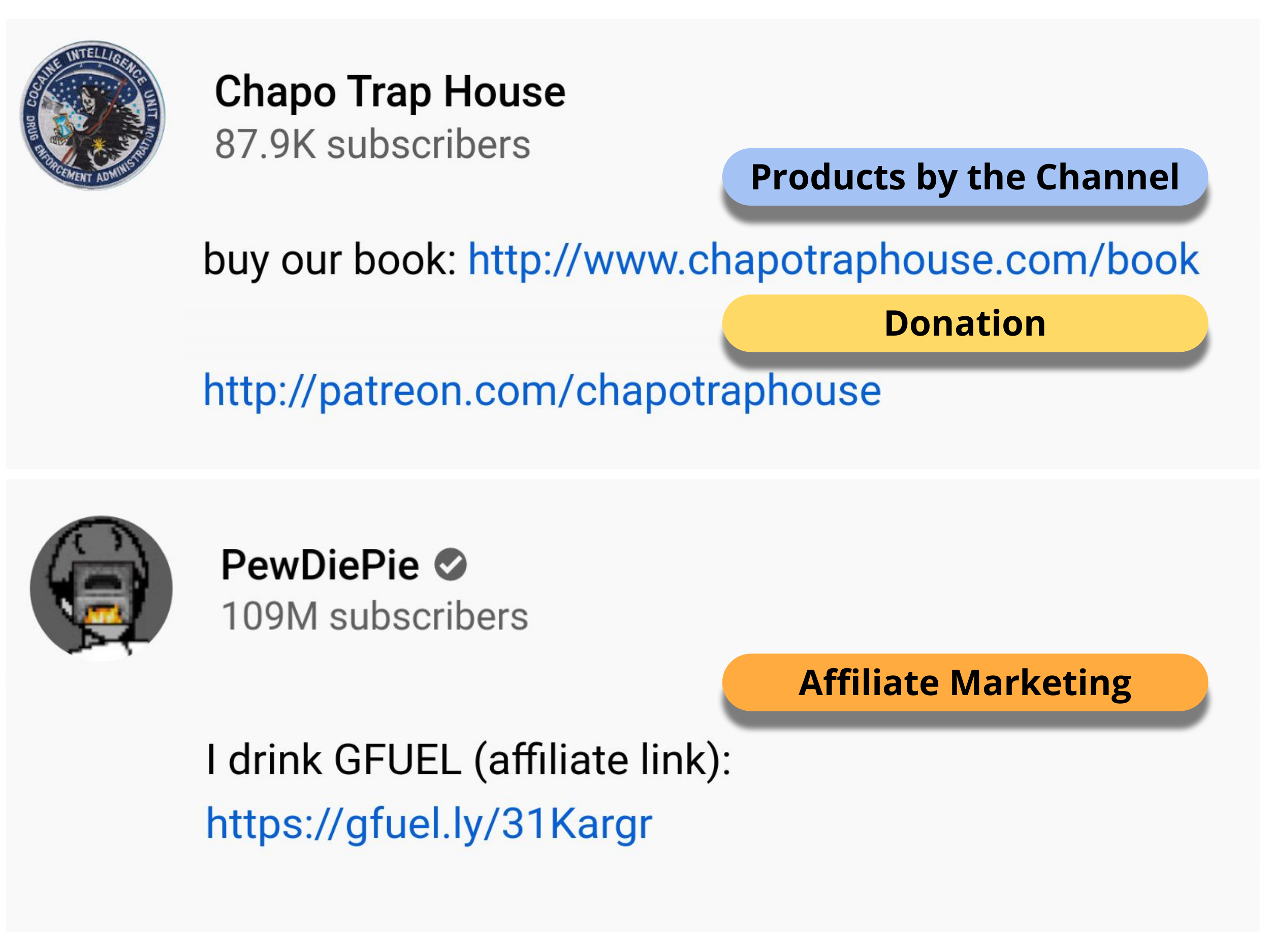}
\caption{}
\end{subfigure}
\hfill
\begin{subfigure}[b]{0.48\textwidth}
\centering
\includegraphics[width=\linewidth]{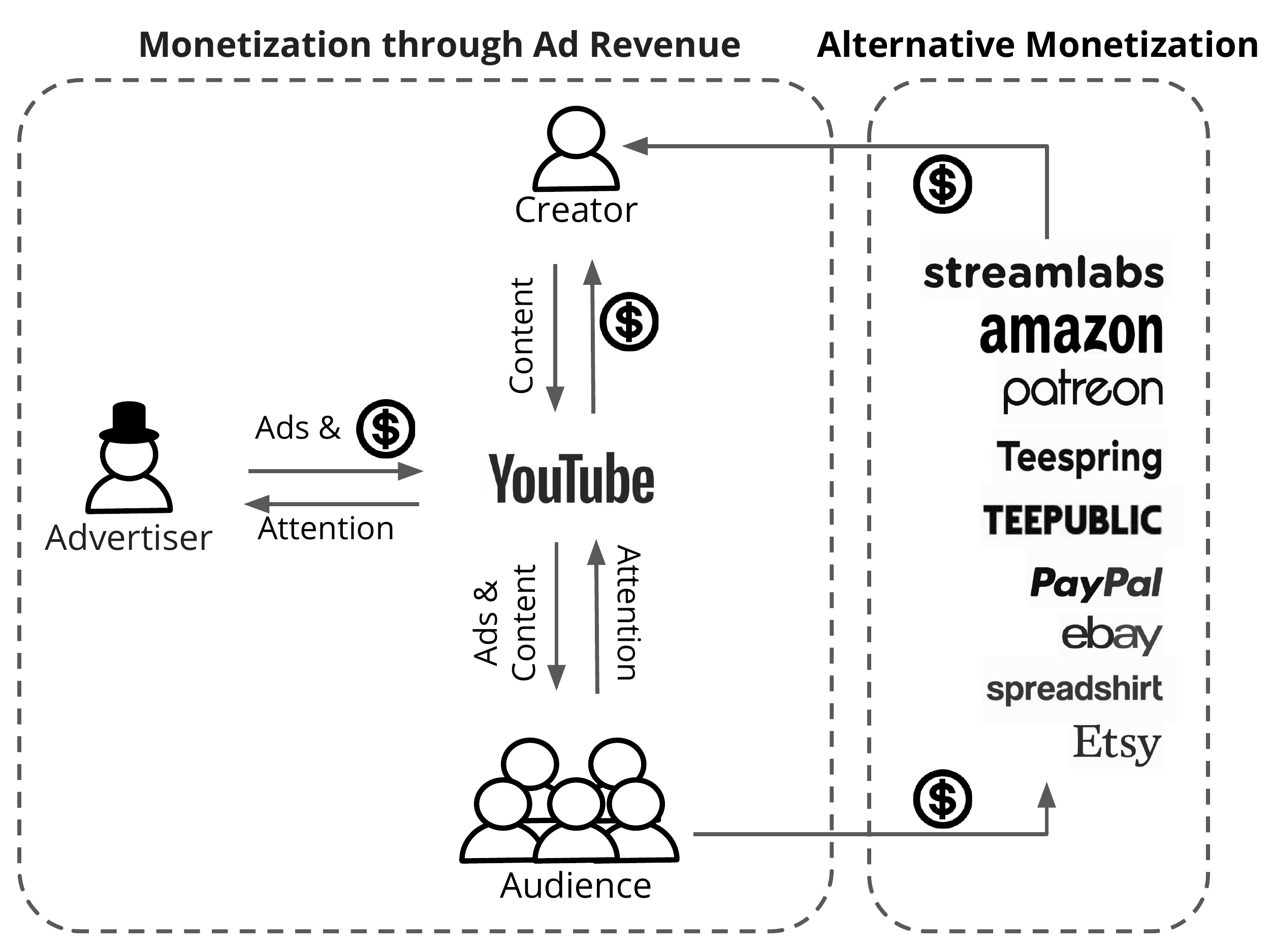}
\caption{}
\end{subfigure}
\caption{\textbf{Alternative monetization strategies:} 
(a) Example video descriptions on YouTube employing alternative monetization practices, labeled (in color) with the type of strategies used in each. 
(b) Schema portraying how alternative monetization strategies allow creators to repurpose YouTube's affordances to earn money directly from their audience.}

\label{fig:opening}
\end{figure}

At the same time, platforms are often co-opted and their affordances are repurposed and exploited by users~\cite{burgess2018youtube,plantin2018infrastructure,nieborg2015crushing,van2015social}. 
One such affordance is the video description, written by the creator and shown alongside YouTube videos.
The description consists of a text box under the video that allows rich formats such as links to other websites. 
Using this text box, creators started including links to alternative monetization services that help them generate revenue outside the YouTube platform. 
For instance, content creators may include a link to a page on \url{patreon.com}, a website that allows direct contributions through one-time payments or recurring subscriptions.
These practices, which we refer to as \emph{alternative monetization strategies}, piggyback on YouTube's content distribution infrastructure to create money streams between viewers and content creators.
Fig.~\ref{fig:opening}(a) illustrates two examples of video descriptions and demonstrates the use of three different alternative monetization strategies. 
The diagram in Fig.~\ref{fig:opening}(b) shows a schematic overview of how alternative monetization contrasts with YouTube's shared ad revenue as a source of the creator's income. 

Alternative monetization may have significant consequences for how YouTube plays its gatekeeper role and for the creators who, as market players, may aim to produce content that maximizes their profits~\cite{munger2020right}. 
For example, creators may no longer care about maximizing exposure, which correlates with ad-based revenue.
Instead, they may emphasize relatability~\cite{marwick2015you} to a smaller but devoted audience who are more likely to contribute.
The consequences of this change in incentives are nontrivial.
On the one hand, weakening the link between exposure and earnings may allow higher-quality content to be produced.
On the other, it may also encourage creators to embrace divisive rhetoric, as observed in Lewis's analysis of reactionary YouTubers~\cite{lewis_this_2020}.
Even if videos are demonetized by YouTube for breaching their policy, it could be that, due to alternative monetization strategies, creators still have substantial financial incentives to create content espousing false, hateful, and divisive narratives. 

It is thus critical to describe and characterize the mechanisms used for alternative monetization on YouTube.
Although past research has examined YouTube's role as a distribution platform~\cite{rieder_mapping_2020, bartl_youtube_2018, figueiredo_tube_2011}, less attention has been given to how its affordances are used to earn money (e.g., ~\cite{mathur_endorsements_2018}).
In that context, we argue that understanding alternative monetization practices is key to deciphering the incentives that shape our online information ecosystem and to better understanding YouTube's role as a gatekeeper.

To this end, the present paper asks the following research questions:

\begin{itemize}
    \item[\textbf{RQ1}] What is the landscape of alternative monetization strategies used by creators to profit outside of the YouTube platform, and how did it evolve?
    \item[\textbf{RQ2}] What is the relationship between content production and the adoption of alternative monetization strategies?
    \item[\textbf{RQ3}] Are alternative monetization strategies used differently by creators of problematic content who might not be allowed to monetize on YouTube directly?  
\end{itemize}

We explore these questions on a longitudinal YouTube dataset with over 71 million videos published by over 136,000 channels~(Section~\ref{sec:data}). 
On YouTube, ``channels'' refer to the public accounts managed by one or more content creators to upload video content.
To analyze this data, we carefully created a taxonomy of alternative monetization strategies, which consists of four broad categories: 
1) requests for donations through third-party platforms (e.g., PayPal, Patreon);
2) affiliate marketing; 
3) sales of products and services related to the channel; and 
4) requests for cryptocurrency donations (Section~\ref{sec:dev}).
Then, we developed a semi-supervised methodology to automatically detect the usage of such strategies in video descriptions (Section~\ref{sec:mes}), which allows quantitative measurements at scale.
Below, we summarize our key findings.

First, content creators included in the analyzed data widely use alternative monetization strategies.
A majority (61\%) of the channels in the studied data use at least one form of alternative monetization in the aforementioned taxonomy, in a total of 18\% of all videos.
These practices have become increasingly more prevalent over the years, dominated by affiliate marketing and requests for direct donations via services like Patreon.
For example, in 2018, around 20\% of the channels linked to external services to collect donations, compared to only 2.7\% in 2008.
The adoption of such strategies is non-uniform: more popular YouTube channels tend to use them more often, and channels tend to adopt strategies tailored to their content. 
For instance, channels producing ``How to \& Style'' content are more likely to use affiliate marketing~(83.\% vs. 51.2\% among the full set of channels),
as they often review clothes, tools, and products.

Second, we show that the adoption of alternative monetization strategies is linked to channel productivity in terms of the rate of content creation.
Although we cannot draw causal conclusions, we show that in the first year after adopting alternative monetization, a channel's content production increases by 43\% on average.
Therefore, when analyzing the content creation dynamics on YouTube, alternative monetization strategies should be considered an important factor contributing to higher productivity.

Third, we show that the practice of using alternative monetization strategies is widespread among problematic content creators, who are likely to be---or at risk of being---shut out from directly monetizing on the YouTube platform. 
Several high-profile incidents brought attention to creators of problematic content making money from their YouTube videos, even when being restricted by YouTube~\cite{coaston_youtube_2018, lunden_youtube_2018, williamson_conspiracy_2018, verge}.
We found that channels from previously studied fringe communities, i.e., the Alt-lite, the Alt-right~\cite{ribeiro_auditing_2020,lewis_alternative_2018}, and the Manosphere~\cite{mamie2021anti}, are more likely to adopt monetization, tend to advertise monetization links more frequently, and use a more diverse set of strategies to monetize, as compared to a carefully chosen comparison set of similar channels.
These fringe channels also rely more on asking for donations through cryptocurrencies or subscription and fundraising services, often employing particular subscription platforms such as \url{subscribestar.com} and \url{hatreon.net}.
As YouTube has taken ``demonetization'' actions on their platform against fringe and other problematic content~\cite{youtube_borderline}, the prevalence of alternative monetization may weaken or even invalidate the effect of this type of platform gatekeeping.

Overall, our results call for more nuance in analyzing the monetary incentives on YouTube and other content-distributing platforms.
This view may help better understand the practices and incentives of content production in the micro-celebrity age~\cite{lewis_this_2020, munger2020right}, as well as the roles of online platforms as curators and gatekeepers~\cite{gillespie2010politics,gillespie2018custodians}.

\section{Background and Related Work}
\label{sec:related}

In this section, we review the background and previous work on YouTube as a platform, its content creation dynamics, and its struggles with problematic content.

\subsection{YouTube as a platform}

Research in the area of platform studies came on the heels of the works on participatory culture~\cite{benkler2006wealth, jenkins2008convergence}. 
In studying online platforms such as YouTube, scholars are interested in the tension between the multiple roles they play: providing ``affordances that support innovation and creativity'' on the one hand, and ``constrain participation and channel it into modes that profit'' on the other~\cite{plantin2018infrastructure}. 
One of these tensions is the conflict between allowing creators to have free and flexible expression, and platforms’ profit goals and their need to define legitimate use~\cite{van2013understanding}.

A significant body of research has documented YouTube as a platform~\cite{gillespie2010politics,plantin2018infrastructure,wyatt2004danger,van2013understanding} and the consequences of viewing it as such.
According to Gillespie, the carefully chosen term ``platform'' allows YouTube to portray itself as a neutral facilitator, instead of a gatekeeper who intervenes in what content gets published~\cite{gillespie2010politics}.
However, as platforms grow larger and become more and more important in our information ecosystem~\cite{burgess2018youtube,van2013understanding},
their choices of what and how content gets published directly impact the public discourse~\cite{gillespie2010politics,van2013understanding,gillespie2018custodians,van2018platform}.

Indeed, YouTube heavily engages in gatekeeping and content moderation actions, determining who can share, gain distribution, and monetize on the platform~\cite{youtube_rec_change,youtube_borderline}. 
YouTube has community guidelines that, when violated, can lead to the removal of videos and the suspension of channels. 
Further, YouTube moderates videos that are allowed on the platform but at the borderline of violating policies, not recommending them to users~\cite{youtube_borderline}. 
This change in YouTube's recommendation mechanism is in reaction to the recent criticism that users are recommended problematic videos~\cite{tufekci_youtube_2018}.
Beyond publishing decisions, YouTube also has the power over who and what gets monetized on its platform~\cite{ytmonetguidelines}, and often resorts to demonetization as a moderation strategy for problematic content on the platform~\cite{markup,williamson_conspiracy_2018}.

While YouTube allows creators to profit from their content using ad revenue-share, often ignored is how users co-opt the platform affordances to monetize via other means. 
To the authors' knowledge, as of January 2022, YouTube does not regulate the usage of alternative monetization on its platform.
In fact, YouTube sometimes partners with alternative monetization services to provide creators with more opportunities to monetize their content~\cite{verge_yt_alt}.
Analyzing alternative monetization practices allows us to complicate our understanding of the tension between the platform and its users, 
and provides insight into profit-driven motivations for content creation on the platform.
This understanding of off-platform monetization can also shed light on YouTube's moderation actions as a platform, including demonetization measures (whose effectiveness is potentially limited by alternative monetization strategies).

\subsection{Content creation dynamics and motivations}

Prior research has argued that the core business of YouTube is its participatory culture~\cite{burgess2018youtube}: content creators on the platform produce videos that attract an audience, and the audience's attention is then sold to advertisers~\cite{ytmonet}.
The resulting revenue from this participation is generally shared by the creators and the platform on the respective platform's terms~\cite{ytmonet,ytmonetguidelines}.
The YouTube platform houses all kinds of creators, ranging from amateur video producers to mainstream media outlets, who are often motivated by different needs~\cite{munger2020right,burgess2008all,jerslev2016media,lewis_this_2020}, including profit. 
Our work, therefore, extends the previous literature on the motivations and practices of YouTube creators.

The development of the relationship between YouTube and content creators on the platform was not without its hiccups.
Content creators have criticized the unpredictability of the platform's monthly revenue and the enforcement of YouTube policies~\cite{caplan_tiered_2020}.
Some creators found that their livelihoods are often threatened by copyright strikes or by changes in the rules that make their content ineligible for monetization~\cite{christin_drama_2021}, particularly affecting content related to groups that are already disadvantaged or underrepresented~\cite{romano_group_2019, markup2}. 
These tensions have been accentuated around incidents referred to by YouTubers as ``adpocalypses''~\cite{kumar2019algorithmic}: sudden changes of guidelines prompted by legal hurdles and boycotts from advertisers and by the deployment of automated systems on YouTube to demonetize content~\cite{christin_drama_2021}.
Aiming to address these issues, some YouTubers have unionized~\cite{union_engadget,union_mit} to negotiate fair treatment.

As the influence of content creators and their dissatisfaction with the ad-based model rose, an ecosystem of services and platforms that allowed other forms of monetization emerged~\cite{regner2020crowdfunding}.
Patreon (\url{patreon.com}) provides a service where fans can support creators with monthly payments in exchange for custom perks.
Teespring (\url{teespring.com}) provides the infrastructure for creators to sell customized merchandise, such as mugs, t-shirts, and hoodies.
Amazon and numerous other e-commerce marketplaces let creators share sales revenue through sponsored links.
As previously mentioned, we refer to monetization using these off-platform services, which ``piggyback'' on YouTube's platform and distribution infrastructure,
as \textit{alternative monetization strategies}.
We note that these strategies are not always perceived as monetization by viewers: Mathur et al.~\cite{mathur_endorsements_2018} studied endorsements on YouTube, 
showing that few creators disclose the endorsements and that users often fail to understand affiliate marketing as advertising.

The alternative monetization strategies we study are intimately linked to micro-celebrity practices on YouTube.
To attract more attention and the monetary benefit that comes with it,
some content creators adopt micro-celebrity~\cite{marwick2015you} and self-branding~\cite{khamis2017self} practices.
Creators exploit the platform's affordances and video-based communications to appear authentic, accessible, and intimate to their audience~\cite{raun_capitalizing_2018}. 
The platform also hands creators who lack strong public identities the resources to self-promote~\cite{khamis2017self}. 
As viewers develop a strong parasocial relationship with content creators who become ``micro-celebrities,'' they are willing to provide direct donations to support the creators' work, buy a personalized mug, or even attend ``real-life'' (paid) events to meet the creators~\cite{abidin2016aren} -- i.e., support the creator via some of the alternative monetization strategies.

\subsection{Problematic content on YouTube}

Researchers have long documented and studied the activities of fringe communities on YouTube, including the distribution and attention they receive on the platform.
As Tufekci argued in her 2018 New York Times article, YouTube's central technological affordance, its recommendation system, might recommend increasingly more ``extreme'' content to its audience, thus arguably making the platform ``the great radicalizer''~\cite{tufekci_youtube_2018}.
The ``radicalization pipeline'' was empirically observed by Ribeiro et al.\ in a large-scale quantitative study, where they showed that users consistently migrated from contrarian communities to overtly white supremacist content~\cite{ribeiro_auditing_2020} (the cause of this pipeline, however, has been largely debated~\cite{munger2020right}).
Although YouTube eventually adopted a more rigid stance towards these channels and banned several of the most prominent ones~\cite{perez_youtube_2020},
a recent report published in 2021 shows that exposure to extremist content remains disturbingly high~\cite{nyhan2021exposure}.
For example, approximately one in ten of that study's participants had viewed at least one video from an extremist channel (9.2\%)~\cite{nyhan2021exposure}.
Beyond far-right content, previous work has also shown other problematic content on YouTube, including harassment~\cite{mariconti2019you} and misogynistic videos~\cite{mamie2021anti,papadamou2020understanding}.

The aforementioned work on the dynamics and motivations behind content creation on YouTube is linked to the rise of problematic content on the platform.
Munger and Phillips~\cite{munger2020right} suggest that the ease of producing, distributing, and monetizing content would lead to a more diverse political media environment, which would include more extreme views. 
According to Lewis~\cite{lewis_this_2020}, political content creators adopt micro-celebrity practices that stress relatability, authenticity, and accountability to appear more credible than mainstream media.
In this work, we show that the content creators who may be more likely to post content that potentially violates YouTube's policies take advantage of alternative monetization strategies. 
This trend is particularly relevant since these creators are often subjected to demonetization~\cite{youtube_borderline}.

\begin{table*}[!t]
    \centering
    \small
    \caption{\textbf{Data summary:} Basic statistics of the datasets used in this paper.}

    \begin{tabular}[t]{llrr}
      \toprule
      \textbf{Dataset} & \textbf{Category} &\textbf{Videos} & \textbf{Channels}\\ \midrule
      YouNiverse  & &71,089,725 & 136,091\\
      \midrule
      &Alt-lite & 78,078 & 114\\
      Fringe &Alt-right & 17,984 & 86\\
      &Manosphere  & 63,191 & 279\\
      \bottomrule
    \end{tabular}
    \label{tab:summary}
\end{table*}

\section{Methods}

To provide a robust view of the alternative monetization strategies used on YouTube, our methods build on YouTube datasets with appropriate channel\hyp{} and video\hyp{}level metadata to
1)~create a taxonomy of alternative monetization methods on the platform; and
2)~develop techniques to automatically identify the types of alternative monetization strategies used by different YouTube channels, extracting information from the URLs appearing in video descriptions.
These methods provide the core input for our analysis of alternative monetization schemes on YouTube.

\subsection{Data}
\label{sec:data}
To address our research questions, we seek to acquire two types of datasets. First, we need a dataset that is generally representative of YouTube channels/videos and that provides a large enough sample to support broad coverage of the trends in the data.
Second, to answer RQ3 about problematic content, we require a dataset that contains data from channels known to produce such content.
Ideally, the datasets would have significant temporal overlap.
To satisfy the first goal of broad coverage, we employ data from the \emph{YouNiverse}~\cite{ribeiro2020youniverse} dataset. 
To allow the analyses of problematic content creators, we compile datasets of ``Alt-right'', ``Alt-lite'' and ``Manosphere'' creators collected by researchers in previous work~\cite{mamie2021anti,ribeiro_auditing_2020}.
We summarize the data in Table~\ref{tab:summary} and describe it below.

\begin{itemize}

\item \textbf{YouNiverse:}
A dataset comprised of 71,089,725 videos from 136,091 English-speaking YouTube channels~\cite{ribeiro2020youniverse}. 
To collect the data, the researchers used websites that collect lists of YouTube channels and the metadata related to the popularity of the channels (number of views and of subscribers) over time.
The researchers then collected video metadata directly from YouTube (in late 2019)~\cite{ribeiro2020youniverse}. 
Channels present in \emph{YouNiverse} were compared with the catalog of \url{socialblade.com}, a prominent tracker of YouTube statistics that tracks over 23 million YouTube channels.%
\footnote{https://socialblade.com/info}
Researchers found that the \emph{YouNiverse} dataset covers around 25\% of the top 100k most popular YouTube channels and around 35\% of the top 10k most popular (both measured by number of subscribers)~\cite{ribeiro2020youniverse}. All channels in the dataset had at least 10,000 subscribers when data collection happened.

\item \textbf{Alt-lite and Alt-right:} 
A dataset comprised of 138,538 videos from 291 YouTube channels~\cite{ribeiro_auditing_2020}.
Each channel was grouped into one of three categories assigned by the same researchers: the Intellectual Dark Web, the Alt-lite (I.D.W.), and the Alt-right. 
The data was collected through snowball sampling: from a set of seed channels,  authors found and annotated other channels through YouTube's recommendation and search functionality~\cite{ribeiro_auditing_2020}.
For our analyses here, we focus on the 200 channels in the dataset labeled by the authors as Alt-lite or Alt-right, two categories containing far-right content creators.

\item \textbf{Manosphere:} A dataset comprised of 63,191 videos from 279 YouTube channels related to Manosphere content~\cite{mamie2021anti}. 
This dataset was collected with the same methodology and during the same period as the ``Alt-lite and Alt-right'' dataset described above. 
The dataset includes data related to a mix of anti-feminist groups broadly referred to as the ``Manosphere,'' an eclectic denomination that includes communities such as Incels, Men's Rights Activists, Men Going Their Own Way, and Pick Up Artists~\cite{lewis_alternative_2018, mamie2021anti}.

\end{itemize}

\noindent
For simplicity, we group together the data from all Alt-lite, Alt-right, and Manosphere channels across the two latter datasets and refer to it as the \emph{Problematic Content} dataset in the rest of this paper.
In the \emph{YouNiverse} dataset, we consider a channel's content category to be the most frequent category assigned by content creators when uploading their videos (e.g., Music, People \& Blogs, News \& Politics).
In the \emph{Problematic Content} dataset, we consider a channel's content category to be the labels assigned by the annotators, categorizing channels as belonging either to the Alt-lite, the Alt-right, or the Manosphere.
Most of the videos in our datasets have view counts that were collected at crawl time.%
\footnote{Between the 12th and the 17th of September 2019 for the \emph{YouNiverse} dataset; Between 19th and 30th of May 2019 for the \emph{Fringe} dataset.}
For videos that did not have a view count~($0.003\%$ of the combined dataset), we impute the average number of views for videos on the respective channel.

\subsection{Developing a taxonomy of alternative monetization methods on YouTube}
\label{sec:dev}
We devised a taxonomy of alternative monetization methods on YouTube  by qualitatively coding a random sample of our datasets.
In what follows, we describe how we obtained this random sample and the details of the coding process.

\xhdr{Obtaining the sample}
We started by sampling videos from the \emph{YouNiverse} and \emph{Problematic Content} datasets.
The sample includes videos with both common and controversial content, 
videos produced at different times, and videos from channels of different popularity and from different content categories.
For each video, we defined three attributes: the number of views it received, the semester in which it was posted (first vs.\ last six months of the year), and its channel content category, which is defined differently in each dataset~(see above).
Considering these attributes, we first split videos into buckets according to their content category and semester.
Then, inside each bucket, we further split videos into five ranges based on the percentiles of their number of views.\footnote{[0, .25], [.25, .5], [.5, .75], [.75, .99], [.99, 1]. Views sorted from small to large.} 
Limited by the capacity of the annotators, for each percentile range, we sampled up to one random video, which resulted in a total of 3,373 videos to be annotated.

\xhdr{Iterative development of the taxonomy}
In the first step of the development of the taxonomy, these 3,373 videos were split between two annotators, both authors of this paper.
Each annotator analyzed the video descriptions
to identify links (e.g., \emph{amazon.com}, \emph{paypal.com}, etc.) 
and keywords (e.g., \emph{donate}, \emph{coupon}, etc.) related to monetization.
The idea here was to carefully document different off-platform monetization strategies used by content creators.
After this initial inspection phase (meant to provide general insight into alternative monetization strategies), annotators shared their notes and developed an initial taxonomy, which was iteratively improved as they re-annotated all 3,373 examples together.
The final taxonomy consists of four different labels, each corresponding to a different alternative monetization strategy on YouTube:

\begin{enumerate}
\item[\textbf{[DO]}] \textbf{Requests for donation}: 
Direct requests for financial support. 
This can be done through a link to a third-party service such as PayPal and Patreon, or to a website owned by the content creator~(e.g., \url{https://www.contentcreator.com/donate/}).
Note that in this category, we also include appeals to contribute to cause-based fundraising platforms such as \emph{GoFundMe}.
In fact, some of the fundraising platforms are known to have been involved in fundraising for far-right causes~\cite{cnbc}. 
We discuss nuances related to this decision in Appendix~\ref{ap:method}.

\item[\textbf{[CR]}] \textbf{Requests for cryptocurrency}: 
Direct donations through cryptocurrencies. 
For example, creators may share their Bitcoin addresses and prompt viewers to send money to them.

\item[\textbf{[PC]}] \textbf{Sales of products/services by the channel}: Selling products and merchandise produced or marketed directly by the creator or the channel.
For example, T-shirts and mugs with the channel's logo on them, or health products that are sold by the creator.

\item[\textbf{[AM]}] \textbf{Affiliate marketing}: Selling products or offering a service not directly associated with the channel or the content creator. Here, creators benefit financially by receiving a referral bonus when a sale is made. For example, creators often link to the gears they use to film their videos on Amazon through a special URL where they get a commission for each sale made. 
\end{enumerate}

Since we employ these categories in subsequent quantitative analysis, we additionally validate the final taxonomy by assessing annotator agreement,
as suggested in the general guidelines proposed for CSCW research~\cite{mcdonald2019reliability}.
The same authors as earlier independently annotated 100~videos with labels according to any of the four alternative monetization practices each video description may contain. 
Considering each category as a binary variable (present/absent), 
Cohen's kappa indicated near-perfect agreement (\textbf{[DO]}:~95\%, \textbf{[CR]}:~100\%, \textbf{[PC]}: 90\%, and \textbf{[AM]}: 89\%).

\subsection{Detecting alternative monetization strategies}
\label{sec:mes}
To explore the alternative monetization ecosystem in our datasets,
we need a scalable method to identify the use of alternative monetization strategies in the videos in our datasets.
For the \textbf{[CR]} category, we use regular expressions to find cryptocurrency addresses.
For the other alternative monetization strategies,
we adopt a similar methodology as in prior work~\cite{mathur_endorsements_2018}.
Specifically, we curate a set of labeled domains related to different alternative monetization strategies and use these to guide our analyses, considering that all instances of the same domain belong to a given monetization strategy.
We detail our method for domain extraction below and release the labeled domains at \emph{[anonymized for review]}.

\xhdr{Identifying Cryptocurrency Requests}
To identify requests for donation through the use of cryptocurrencies (\textbf{[CR]} category), we use a cryptocurrency address validator\footnote{\url{https://github.com/k4m4/cryptaddress-validator}} to extract valid cryptocurrency addresses from the video descriptions.
We match addresses for seven popular coins: Bitcoin, Ethereum, Litecoin, Dash, Ripple, Dogecoin, and Neo.
Additionally, for Bitcoin and Ethereum addresses, we obtain the full transaction history of the corresponding coin addresses using the blockchair API,%
\footnote{\url{https://blockchair.com/}}
which allows us to calculate the lifetime earning for any Bitcoin and Ethereum addresses.

\xhdr{Identifying Use of Strategies Using Associated Domains}
Detecting the use of other types of alternative monetization strategies is less trivial, since there is no exact regular expression to match a video description and tell us whether an alternative monetization strategy is being used. Fortunately, one of the important aspects of alternative monetization strategies is that they have to redirect the user to a third-party website, where a sale or donation will occur. 
Therefore, we resort to a methodology based on label propagation that associates domains present in video descriptions with a specific monetization strategy.

Our methodology is based upon a simple observation. 
Frequently, on YouTube, links are preceded by brief descriptions.
For instance, to display a link selling channel-specific merchandise (\textbf{[PC]}), a content creator may write: 
``\wordformat{buy merch}:
\domainformat{www.teespring.com/YouTuberX}.'' 
These brief descriptions can be abstracted as a sequence of word--domain co-occurrences:
(\wordformat{buy}, \domainformat{teespring}),
(\wordformat{merch}, \domainformat{teespring}). 
Using this abstraction, we can find other category-related words and domains in the data with this single seed example.
For instance, we can look for other words that occur with the domain \domainformat{teespring}, and may find another video description: ``\wordformat{mugs/t-shirts}: \domainformat{www.teespring.com/YouTuberY},''
suggesting that the words \wordformat{mugs} and \wordformat{t-shirts} are also related to the merchandise category (\textbf{[PC]}).
In a similar fashion, to find other domains selling merchandise, we may look for domains that co-occur with the word \wordformat{merch}. 

Leveraging this intuition, we apply the following method to extract domains related to alternative monetization strategies. 
First, we build an undirected bipartite graph representing words and domains. 
For each word--domain pair $(i, j)$ where the word appears before the domain for exactly $k$ times in our datasets, we add an edge with weight $k$ between the nodes corresponding to word $i$ and domain $j$.
Next, we define a set of seeds in this graph: words and domains related to one of the monetization categories, as well as words and domains that are not directly related to monetization. 
For example, seeds for the donation category include words and domains such as \wordformat{donate} and \domainformat{patreon}. The non-monetization category includes words and domains such as \wordformat{follow} and \domainformat{twitter}.
Last, we run a standard label propagation algorithm~\cite{zhu_introduction_2009} on this graph using the aforementioned seed words and domains as the labeled examples.
The algorithm iteratively propagates the label of the seed words and domains to other words and domains that are frequently used together. 
We detail this label propagation methodology as well as how we validate its output in Appendix~\ref{ap:method}.

The output of our method is a set of labeled keywords and domains, each associated with at most one alternative\hyp monetization label.
We only use domains---and not keywords---for the following analyses,
since keywords could also appear in other contexts beyond descriptions that precede external links, which may lead to inaccurate label inferences.
We leverage this set of domains along with the cryptocurrency addresses to infer the types of alternative monetization strategies used by each channel and video in our dataset.

\subsection{Limitations}
We highlight some of the limitations of our methodology and data.
First, our datasets and analyses are limited to English content only.
Second, the datasets we employ were crawled in 2019 and may not include newer types or trends in alternative monetization strategies.
Third, in this initial work, we focus on three types of problematic content, Alt-lite, Alt-right, and Manosphere,
due to their negative impact on the audience and potentially on the society at large.
We leave the exploration on other types of problematic content to future work.
Fourth, our taxonomy was developed on a limited number of samples. 
It is possible that there are alternative monetization strategies which are not included in our sample, and hence are not captured by our taxonomy.
However, we did not observe such a strategy as we analyzed more data manually for 
further analyses.
Such a strategy is also likely not widely adopted, therefore is out of the scope of this initial work. 
Fifth, for the \emph{YouNiverse} dataset, we use the content categories assigned by content creators as a proxy to estimate the content type of channels and video. However, content creators might not share a consistent definition of categories and may mislabel their content, adding noise to our analysis and findings. 
Sixth, we assumed that the video description, collected when each dataset was crawled, had been posted together with the video. In reality, content creators may retrospectively edit previously uploaded videos to include alternative monetization links,  which may affect some of the results of the longitudinal analyses of Section~\ref{sec:longitudinal}.

Lastly, we label monetization practices at the domain level, which may be incomplete or inaccurate in some cases. For instance, creators may share links to merchandise of other creators, or link to products on Amazon or other e-commerce platforms that are not affiliate links.
Also, a single domain may be associated with multiple monetization strategies.
We alleviate concerns associated with this limitation by performing an extensive set of validation experiments (described in Appendix~\ref{ap:method}).
We find that our method is inaccurate in labeling donation-related domains (\textbf{[DO]}), returning many false positives.
This issue was resolved by manually annotating the (relatively few) domains predicted to be donation-related.
After this post-processing step,
for each alternative monetization category, 
we manually inspect 100 URL samples that are labeled with the category, along with the surrounding context of the URL.
We show that $91.5\%$ of the 400 inspected URLs are correctly labeled by our method.
The category with the lowest accuracy is Affiliate Marketing~(\textbf{[AM]}), with 75\% accuracy.
We note that 72\% of the false positive cases for this category are due to the fact that URLs selling channel-related products are misclassified as affiliated marketing.
One typical example is an Amazon link~(a domain often used in the context of affiliate marketing) that sells a book written by the content creator.
These accuracy issues may lead to an over-estimation of the \textbf{[AM]} category and an under-estimation of the \textbf{[PC]} category. 
We argue that this has minimal impact on our main takeaways, as they do not build upon the precise differentiation between these two categories.
\begin{table}[]
\small
\centering
\caption{\textbf{Prevalence of alternative monetization strategies:} 
We depict the prevalence of alternative monetization strategies in the \emph{YouNiverse} dataset, relative to the number of videos (in \textit{a}) and of channels (in \textit{b}).
Both tables show results stratified by 8 of the 15 YouTube content categories as well as aggregated across all 15 categories (\textbf{All}). 
Columns in both \textit{(a)} and \textit{(b)} show results stratified according to the different monetization categories and aggregated across all categories (\textbf{Any}). 
We mark in bold the categories of which the videos or the channels are 1.2 times more likely to use a certain type of alternative monetization strategy,
i.e., the percentage of the entry is 20\% larger than the corresponding aggregated values in the last row.
}
\label{tab:prev}

\begin{tabular}{l|rrrrrr|rrrrrr}

\toprule
& \multicolumn{6}{c}{(a) video level (\%)} & \multicolumn{6}{c}{(b) channel level (\%)} \\
\midrule
Categories & \textbf{PC} & \textbf{AM} & \textbf{DO} & \textbf{CR} & \textbf{Any} & \tiny{n$^{(\times10^5})$}  & \textbf{PC} & \textbf{AM} & \textbf{DO} & \textbf{CR} & \textbf{Any} & {\tiny n $^{(\times10^3})$} \\
\midrule
Entertainment & 2.9 & 8.2 & 6.7 & 0.3 & 14.1 & 117 &  12.8 & 46.0 & 27.3 & 0.6 & 55.7 & 23 \\
Films & 1.7 & 8.8 & 6.4 & 0.1 & 14.3 & 23 & 8.6 & 37.2 & 27.9 & 0.3 & 51.0 & 7  \\
Gaming & \textbf{7.4} & 11.3 & \textbf{18.1} & 0.1 & \textbf{29.4}  & 133 & \textbf{18.3}  &47.5 & \textbf{49.7} & 0.6 & 68.0 & 20 \\
Howto\&Style & 1.5 & \textbf{36.7} & 6.0 & 0.2 & \textbf{39.6} &  38 & 10.0 & \textbf{83.5} & 23.5 & 0.5 & \textbf{85.4} & 12  \\
Music & 1.3 & 7.1 & 4.7 & 0 & 11.5 & 80 & 6.5  &  40.0 & 17.2 & 0.2 & 47.6 & 24  \\
People\&Blogs & 2.7 & 14.4 & 7.3 & 0.3 & 20.2 & 67 &  10.4 & 50.4 & 23.7 & 0.7 & 56.7 & 18 \\
Science\&Tech & 1.7 & \textbf{25.9} & 7.3 & \textbf{0.8} & \textbf{29.3} & 23 & 8.9 & \textbf{77.8} & 33.8 & \textbf{3.2 }& \textbf{79.9}  & 5 \\
Travel & 2.2 & 12.6 & 7.7 & 0.1 & 17.3 & 11   & 11.2 & 57.8 & 30.8 & 0.7 & 62.9 & 2  \\
\midrule 
\textbf{All} & \textbf{3.0} & \textbf{11.3} &  \textbf{8.1} & \textbf{0.2} & \textbf{18.4}  & 694 & \textbf{11.3} & \textbf{51.2} & \textbf{28.4} & \textbf{0.7} & \textbf{60.6} & 136 \\
\bottomrule
\end{tabular}
\label{tab:my_label}
\end{table}

\section{The Prevalence of Alternative Monetization Strategies}
\label{sec:prev}

We begin by broadly characterizing the adoption of alternative monetization strategies in the \emph{YouNiverse} dataset, as shown in Table~\ref{tab:prev}. Recall that here we are considering relatively popular channels (10,000+ subscribers at the time of data collection).
More specifically, we measure the percentage of videos and channels in the \emph{YouNiverse} dataset that contain at least one video description that has a monetization-related domain or cryptocurrency address.

We find that, according to this criterion, 18.4\% of the videos and 60.6\% of the channels in the dataset have used at least one alternative monetization strategy, as shown in the bottom row of Table~\ref{tab:prev}. 
The most common alternative monetization strategy is affiliate marketing \textbf{[AM]} (used in 11.3\% of the videos and 51.2\% of the channels), followed by donation \textbf{[DO]} (8.1\%; 28.4\%), products by the channel \textbf{[PC]} (3.0\%; 11.3\%) and cryptocurrencies \textbf{[CR]}  (0.2\%; 0.7\%). 
These results suggest that alternative monetization practices are extremely prevalent in the YouTube content creation ecosystem.

\begin{figure}
    \centering
    \includegraphics[width=\linewidth]{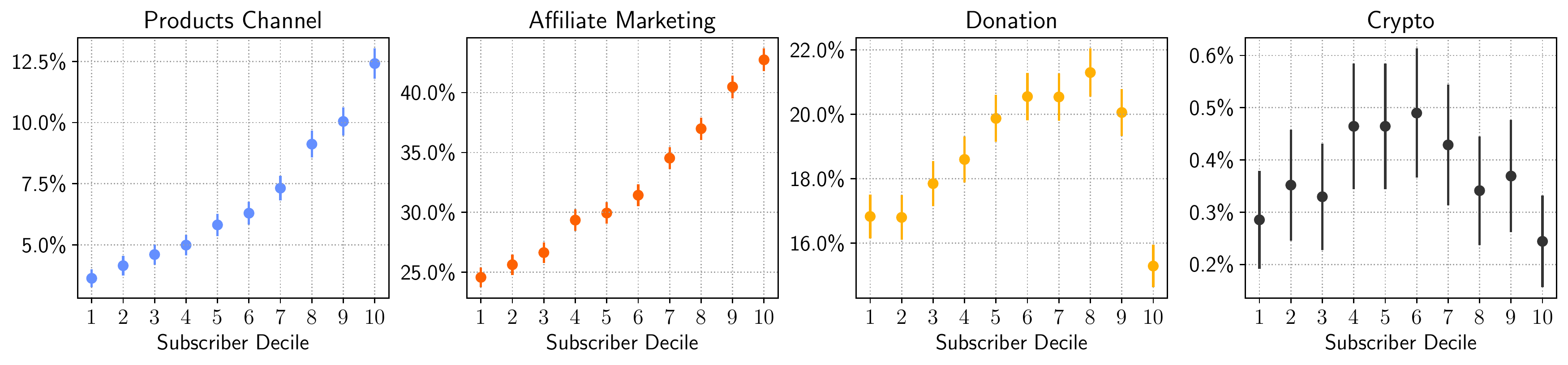}
    \caption{\textbf{Popularity and alternative monetization:} We analyze the relationship between the popularity of channels and the usage of alternative monetization strategies. We divide all channels that have published any video in 2019 according to their number of subscribers in 10 deciles. Then, for each monetization strategy, we calculate the percentage of channels that monetize in each decile. Note that the $y$-axis scales differ and that error bars represent 95\% CIs. We repeat the analyses done in this figure dividing channels in quintiles and vigintiles (i.e., groups of 5 and of 20), and obtained largely the same results.}
    \label{fig:subs}
\end{figure}

Additionally, we also find that different video and channel categories have different monetization ``fingerprints,'' that is, their choice of alternative monetization strategies varies greatly.
This can be observed in the adoption percentages stratified per category, also in Table~\ref{tab:prev} (we depict eight out of the 15 content categories available on YouTube, the same used by Mathur et al.~\cite{mathur_endorsements_2018}).
For example, \textit{How To \& Style} videos often review clothes, tools, and products. In that context, it is not surprising that we find that channels and videos in this category use the affiliate marketing \textbf{[AM]} monetization strategy much more than the general percentage~(36.7\% of videos; 83.5\% of channels).
Gaming channels, on the contrary, use the affiliate marketing \textbf{[AM]} monetization strategy less often (11.3\%; 47.5\%), but are much more likely to sell merchandise (7.4\%; 18.3\%) and ask for donations (18.1\%; 49.7\%).
Altogether, these results suggest that content creators tailor their alternative monetization strategies to the content that they produce.

We compare our results with the work of Mathur et al.~\cite{mathur_endorsements_2018} on affiliate marketing on YouTube.
According to their measurements, for instance, 3.49\% of \textit{How To \& Style} videos present affiliate marketing links (compared to 36.7\% in our measurement). 
This large difference between our results can be explained in two ways.
First, the sampling strategies used by their paper and by the \emph{YouNiverse} dataset are very different. They use a random prefix sampling, which yields a non-biased sample of YouTube videos, whereas \emph{YouNiverse} focuses on large channels with more than 10,000 subscribers.
Hence, the \emph{YouNiverse} dataset contains more popular channels than a non-biased sample.
As we show later in Figure \ref{fig:subs}, channels that are more popular are more likely to use alternative monetization strategies.
Second, their methodology is much more specific: tracking the URLs of specific companies that contain some sort of affiliate identifier.
Our analysis suggests that the lack of effective disclosure of endorsements exposed previously may have been underestimated.

We further explore the relationship between channel popularity and alternative monetization strategies in Figure~\ref{fig:subs}, showing that generally speaking, popularity is positively correlated with the adoption of alternative monetization strategies. 
We operationalize popularity via the number of subscribers that the channels had at the time of the collection of the \emph{YouNiverse} dataset (which, as previously discussed, only contains channels with over 10,000 subscribers). 
Then, we divide channels into 10 deciles based on their number of subscribers.\footnote{Deciles: 12.7k, 16.4k, 21.8k, 29.8k, 42.2k, 63k, 102k, 185k, 440k.}
In the figure, for each decile ($x$-axis), the $y$-axis specifies the percentage of channels in the decile that use the respective monetization technique at least once.
For the \textit{Products Channel} \textbf{[PC]}, and the \textit{Affiliate Marketing} \textbf{[AM]} categories, the figure shows a clear increasing trend: the more popular the channel, the more likely it is to monetize in those ways.
For example, 12.5\% of channels in the 10th decile (with more than 440k followers) have posted a link selling merchandise (category \textit{Product Channel}), compared to 3.6\% of channels in the first decile (those with less than 12.7k followers).
The relationship between channel popularity and the usage of cryptocurrency-based monetization (\textbf{[CR]}) is much noisier. Channels in the top and the lower deciles share cryptocurrency addresses less than channels in the middle deciles.
Interestingly, for the \textit{Donation} category (\textbf{[DO]}), we observe a drop in the percentage of channels that monetize for the last two deciles.
This is likely because some high-profile channels, for example, those from famous artists or late-night talk shows, do not need to rely on direct donations from fans.
We repeat the analysis on the relationship between channel popularity and alternative monetization strategies stratified by category in Figure~\ref{fig:subs_cats} (placed at the end of this paper). Overall, we observe the same trend across the different categories: more popular channels use alternative monetization more frequently.

Lastly, in Table~\ref{tab:major_domains}, we study the prevalence of the top monetization domains in the dataset, analyzing the top three most popular domains in each alternative monetization strategy.
We report 
1) the \textit{number of occurrences} of each domain, i.e., the number of times that the domain is used in any video description; 
2) the channel level prevalence of each domain; and lastly,
3) the \textit{employment rate} of each domain, i.e.,  the chance of a specific domain being employed when a given alternative monetization strategy is used.
For cryptocurrencies, we report analogous results for the top 3 most popular coins: Bitcoin (BTC), Litecoin (LTC), and Ethereum (ETH).
We find that alternative monetization strategies are surprisingly concentrated in the hands of a few players.
For example, \domainformat{amazon}, the most popular domain overall,  is linked by 32.0\% of channels and corresponds to 44.0\% of all usages of any \emph{Affiliate Marketing} \textbf{[AM]} domain.
Also, in 92\% of the times content creators use a \emph{Donation} \textbf{[DO]} domain to monetize,
they resort to one of the three top services: \domainformat{patreon}, \domainformat{paypal} and \domainformat{streamlabs}.
These results show that within individual alternative monetization strategies, the usage is heavily concentrated in the hands of a few players.

\begin{table}[]
\small
\centering
\caption{\textbf{Prevalence of top domains/coins for each alternative monetization strategy:} For the top three most popular domains in each monetization strategy, we show their number of occurrences, the percentage of channels that have linked to that domain, and, lastly, the employment rate of the domain (so, for instance, of all usages of Affiliate Marketing, 44.0\% of the times the creator links to Amazon). Following the same logic, we also show the prevalence of cryptocurrency addresses for Bitcoin (BTC), Litecoin (LTC), and Ethereum (ETH). Here, instead of calculating the number and the percentage of all URLs, we calculate the number of addresses and the percentage of all crypto addresses.}

\begin{tabular}{r|ccc|ccc}
\toprule
& \multicolumn{3}{c}{\textbf{[PC]}} & \multicolumn{3}{c}{\textbf{[AM]}}\\ \midrule 
& \domainformattable{teespring} &  \domainformattable{spreadshirt} &  \domainformattable{teepublic} &  \domainformattable{amazon} &  \domainformattable{ebay} &  \domainformattable{etsy}\\
\# Occurrences $^{\times 10^4}$ & 89 & 107 & 9 & 1562 & 35 & 47\\
\% Channels & 5.7\% & 2.7\% & 0.4\% & 32.0\% & 5.9\% & 5.2\%\\
Employment Rate &  30.9\% & 37.1\% & 3.1\% & 44.0\% & 1.0\% & 1.3\%\\ \midrule
 & \multicolumn{3}{c}{\textbf{[DO]}} & \multicolumn{3}{c}{\textbf{[CR]}} \\
  &  \domainformattable{patreon} &  \domainformattable{paypal} &  \domainformattable{streamlabs} &  \domainformattable{BTC} &  \domainformattable{LTC} & \domainformattable{ETH} \\ 
\#Occurrences $^{\times 10^4}$ & 491 & 217 & 95 & 16.3 & 8.4 & 7.2\\
 \% Channels  & 16.6\% & 7.8\% & 5.3\% & 0.67\% & 0.24\% & 0.35\% \\ 
Employment Rate  & 56.3\% & 24.9\% & 11.0\% &  46.7\% &  24.3\% & 20.6\% \\
 \bottomrule
\end{tabular}
\label{tab:major_domains}
\end{table}

\section{Evolution of Alternative Monetization Strategies over Time}
\label{sec:longitudinal}

The above analyses (Section~\ref{sec:prev}) did not capture how the usage of alternative monetization strategies changes over time.
To further explore this aspect, we now perform an additional set of longitudinal analyses on the usage of alternative monetization on the \emph{YouNiverse} dataset.

\begin{figure}
\centering
\begin{subfigure}[b]{0.48\textwidth}
\centering
\includegraphics[width=0.97\linewidth]{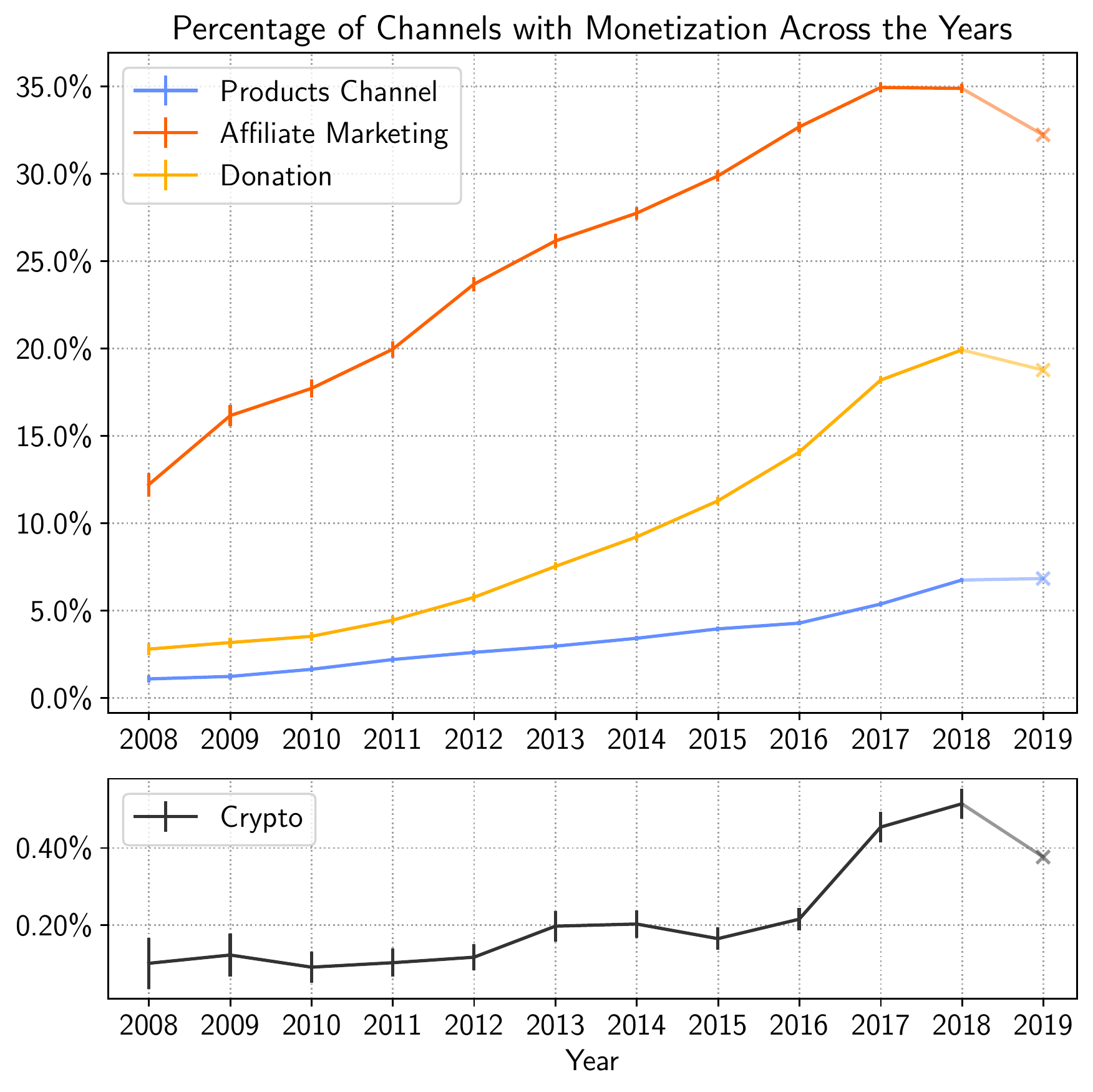}
\caption{}
\end{subfigure}
\hfill
\begin{subfigure}[b]{0.48\textwidth}
\centering
\includegraphics[width=0.97\linewidth]{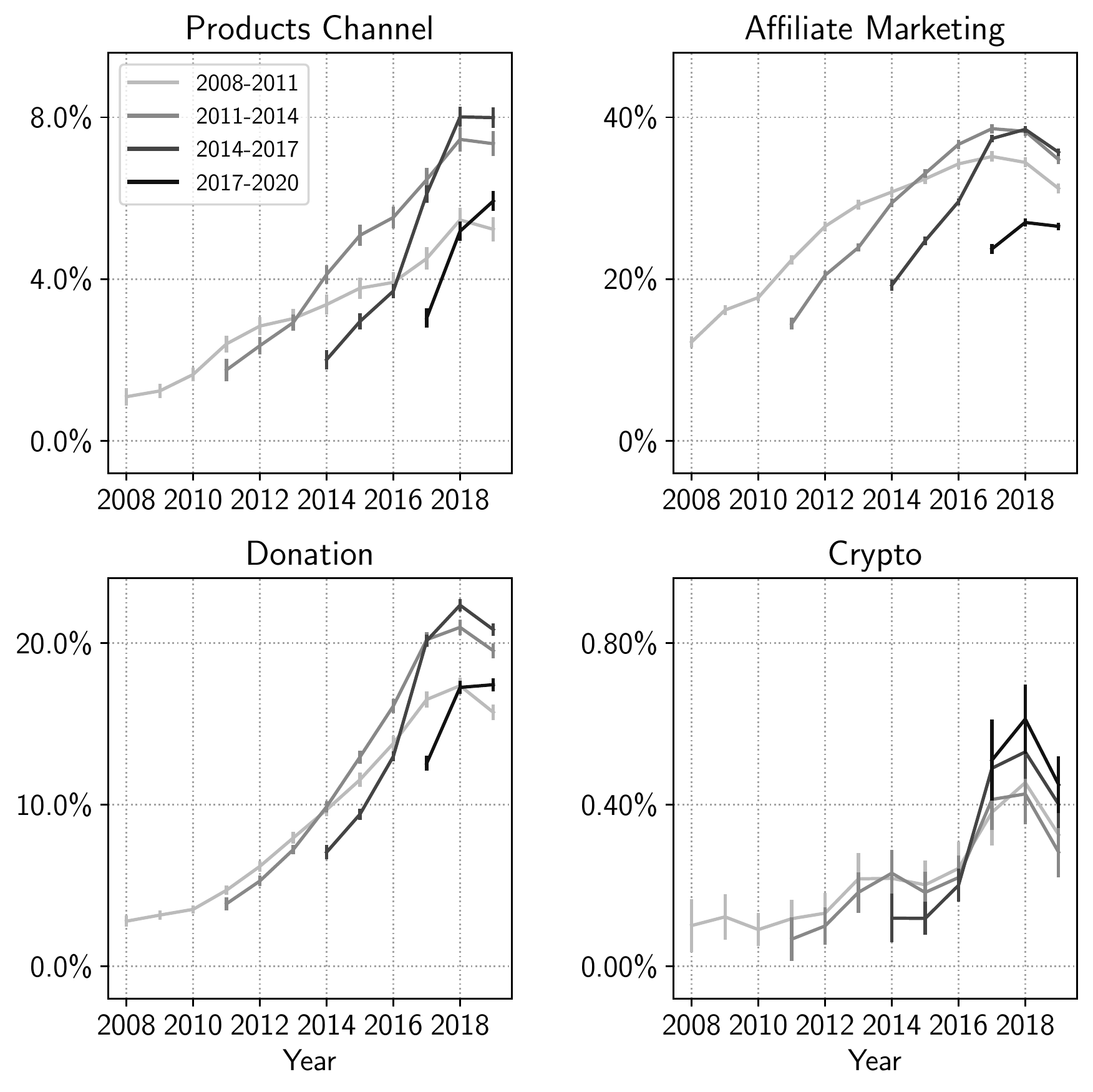}
\caption{}
\end{subfigure}
\caption{\textbf{Channel-level usage of alternative monetization strategies:} (a) We show the prevalence of monetization strategies across the years in the \emph{YouNiverse} dataset. We consider that a channel uses a monetization strategy in a given year if its videos have at least one URL associated with the strategy or contains a cryptocurrency address.
(b) We analyze the relationship between the age of channels and the usage of alternative monetization strategies. We divide channels into four cohorts according to their creation date. 
Then, for each monetization strategy, we calculate the percentage of channels in each cohort that used this strategy in a given year. Error bars represent 95\% CIs.}
\label{fig:channel_adoption}
\end{figure}

Figure~\ref{fig:channel_adoption}(a) depicts the usage of alternative monetization strategies across the years on the channel level.
The $x$-axis specifies the years, from 2008 to 2019, and the $y$-axis shows the percentage of channels that used each alternative monetization strategy (shown in different colors) in a given year.
We consider that a channel has used an alternative monetization strategy if at least one of their videos that year contains one of the domains associated with that monetization strategy (via our semi-supervised methodology described in Section \ref{sec:data}). 
Note that the year 2019 is incomplete in the dataset employed,
as \emph{YouNiverse} data was collected around August 2019~\cite{ribeiro2020youniverse}, which explains the decline in the plot.
Here, we also show the yearly fraction of channels that post at least one video with a cryptocurrency address in its description.

We find that among the channels in the \emph{YouNiverse} dataset, the usage of alternative monetization strategies has been growing steadily.
In 2019, 32.2\% of channels had at least one URL linking to a domain in the category \textit{Affiliate Marketing} (vs. 12.2\% in 2008), 18.7\% posted at least one URL with a \textit{Donation} domain (vs. 2.7\% in 2008), and 6.8\% posted at least one URL linking to \textit{Product Channels} (vs. 1.1\% in 2008). 
Cryptocurrencies are by far the least used of all alternative monetization strategies: in 2019, only around 0.4\% of channels had a video with a cryptocurrency address (vs. 0.1\% in 2008).
These results should be treated as an upper bound, since channels can edit their video descriptions retrospectively to add URLs with monetization-related domains or cryptocurrency addresses.
However, regardless, we argue that the rising trend in the usage of all four alternative monetization strategies suggests that YouTube's content creation ecosystem is increasingly reliant on the money brought forth by donations, affiliate marketing, and merchandising.
In Figure~\ref{fig:prev_cats} we repeat the analysis done in Figure~\ref{fig:channel_adoption}(a), but considering distinct channels categories separately. Again, as in Section~\ref{sec:prev}, we find that different categories have very different monetization ``fingerprints,'' but the adoption of alternative monetization strategies is growing across all categories.

The previous analysis does not distinguish between changes in the entire pool of channels producing content for YouTube and changes in behaviors of a fixed set of channels. 
To disentangle these two components, we conduct an additional cohort-based analysis. 
We split channels into four cohorts depending on the date of their first uploaded video,%
\footnote{Cohorts: $[2008,2011)$, $[2011,2014)$, $[2014,2017)$, $[2017,2019)$)}
and analyze their usage of alternative monetization strategies separately.
Results are shown in Figure~\ref{fig:channel_adoption}(b), where, for the different cohorts (represented by lines), we measure the channel-level prevalence~($y$-axis) for each of the four alternative monetization strategies over the years~($x$-axis).

We find that the usage of alternative monetization strategies is consistent throughout the different cohorts: alternative monetization is becoming more prevalent for both older and newer channels.
For instance, in 2019,  around 35\% of the channels in both the $[2011,2014)$ and the $[2014,2017)$ cohort used the Affiliate Marketing \textbf{[AM]} strategy.
This suggests that the increase in the prevalence of alternative monetization strategies observed in Figure~\ref{fig:channel_adoption}(a) is not happening due to the replacement of old channels (which do not use these strategies) by newer channels (which do).

Although the trend is similar for both newer and older channels, different cohorts still differ systematically. For example, we find that younger channels start to use alternative monetization strategies relatively faster, as evident in the higher starting points of the individual cohort curves.
For example, in 2017, the very first year of the $[2017,2019)$ cohort, 27.0\% of the channels in the $[2017,2019)$ cohort used \textit{Affiliate Marketing} monetization strategy.
In comparison, for the $[2014,2017)$ and $[2011,2014)$ cohorts, only 19.2\% and 14.4\% used this strategy in their first years (2014 and 2011, respectively).

One particularity of the analyses done in this section is that the data considered was collected in late 2019~\cite{ribeiro2020youniverse}, and includes only channels with over 10,000 subscribers at the time of the data collection.
As YouTube has grown substantially over the period analyzed, the criteria for the inclusion of data is stricter for earlier dates (for instance, having 10,000 subscribers in 2008 was likely harder than in 2018). 
In that sense, our analysis may overlook channels that were relatively popular in earlier years, but that never managed to reach 10,000 subscribers by 2019.
Moreover, \emph{YouNiverse} data has a "popularity bias," covering YouTube channels with many subscribers more extensively. 
Therefore, it could be that channels that adopt alternative monetization become more popular and thus are more likely to be included in the dataset.
\section{Relation between content creation and usage of alternative monetization strategies}
\label{sec:longevity}
\begin{figure}
    \centering
    \includegraphics[width=\linewidth]{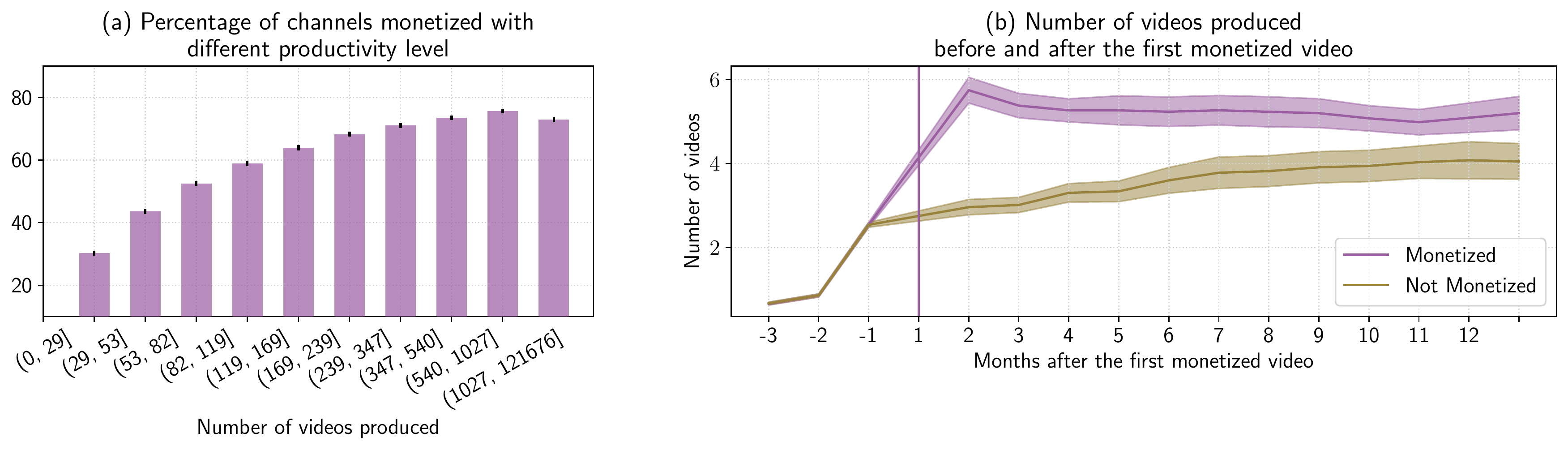}
    \caption{\label{fig:longevity}
    \textbf{Alternative monetization and channel productivity:} (a) Percentage of channels that have alternative monetization link in at least one of their videos, with different productivity levels. 
    The channels are grouped into ten deciles, according to the number of videos they produced.
    Error bars represent 95\% CIs for binomial proportions.
    (b) Comparison of channel productivity between channels that adopted alternative monetization and those that did not. Month 0 specifies the month when one of the groups started alternative monetization.
    Error bars represent 95\% CIs.
    }
\end{figure}

In Section~\ref{sec:prev}, we have shown that more popular channels, i.e., with more subscribers, are more likely to adopt alternative monetization strategies (cf.\ Figure~\ref{fig:subs}).
Undoubtedly, alternative monetization adds an extra monetary incentive to create content
and has the potential to impact a channel's growth and success.
In this section, we analyze the relationship between alternative monetization and content creation.
First, we broadly characterize how the productivity of a channel relates to the likelihood of it adopting an alternative monetization strategy.
Then, we gear the second analysis towards understanding whether the adoption of an alternative monetization strategy at a given point in time indicates an increase in content productivity afterward.

Overall, channels that produce more content are more likely to adopt alternative monetization.
Figure~\ref{fig:longevity}(a) shows the percentage of channels adopting alternative monetization strategies~($y$-axis) stratified by productivity levels~($x$-axis).
From left to right, 
channels are grouped into 10 deciles, according to the number of videos posted.
For example, 
among the least productive $10\%$ of the channels~(left-most bar), who produced at most 29 videos,
only $30\%$ adopt alternative monetization.
The adoption rate is much higher among more prolific channels.
Over $72\%$ of the most productive $10\%$ of the channels~(right-most bar) include alternative monetization links in their video descriptions.
This trend remains the same when channels are grouped into quintiles or vigintiles.
Note that this trend may also be explained by the fact that channels that started earlier are more likely to adopt alternative monetization, and they are more likely to have produced more videos.

The above result shows that a channel's productivity correlates with the adoption of alternative monetization strategies.
However, our findings do not distinguish between two competing hypotheses. 
It could be that 1) channels that adopt alternative monetization strategies become more productive, or that 2) channels that are already productive and popular are the ones who begin to use alternative monetization strategies.
To tease the two hypotheses apart, we conduct an additional study where we compare the productivity (measured in number of videos) between matched pairs of channels. In each pair,
one channel adopted an alternative monetization strategy for the first time in a given month $x$, and the other has not adopted any alternative monetization strategy until month $x$, where $x$ is a month that ranges between January 2015 to September 2019.
Channels are matched according to their content, productivity level, and the amount of attention received in the three months leading to month $x$ (when one of the channels adopted an alternative monetization strategy).
More specifically, for each month, we stratify channels in ten deciles according to their number of views and ten deciles according to the number of videos they produced (yielding a combination of 100 possible combinations per month, e.g., in month $x-1$ a channel may be in the second view-related decile, and the third video-related decile).
Then, we match two channels, if and only if, they are of the same content category, 
and for each of the three consecutive months before month $x$, they are in the same decile in terms of both number of videos produced and number of views received from those videos.
We do not consider channels that have produced no video or received no view in a given month, and discard the top 5\% of the channels that produced the most amount of videos and received the largest number of views (to exclude outliers).
In total, we obtain 5,730 pairs of matched channels.

Figure~\ref{fig:longevity}(b) shows the number of videos produced by two groups of channels,
channels that adopted alternative monetization strategies in month 0~(in magenta) and channels that did not~(in dark yellow).
Month 0, where one group adopts monetization, is marked by a vertical line.
The $x$-axis specifies the number of months from month 0.
Negative values represent months before the monetization happens,
the period when the two groups are matched.
Positive numbers represent months after
one of the matched channels adopted an alternative monetization strategy.
The $y$-axis specifies the average number of videos produced by the channels in each given month.
Note that the two groups are matched on their activity before $x=0$~(left to the vertical line). During this period, the two lines overlap, showing that the matching was effective.
The effect is similar across all four types of alternative monetization strategies.

We find that, 
for the channels studied,
the adoption of alternative monetization strategies has a significant positive correlation with the productivity of a channel, both in the short and long term.
For example, during the first month, after one of the channels adopted an alternative monetization strategy~($x=1$), we already see a large difference between the productivity of channels. Channels that adopted an alternative monetization strategy in month $x=0$ on average produced six videos, while the channels that did not, only produced three.
The gap becomes narrower with time, yet the difference remains.
Over the entire year, aggregating the number of videos produced between months 1 to 12, channels adopting alternative monetization strategies produced 63 videos on average while those who did not produced 44 videos.
This amounts to an increase in productivity of over 43\%.

We stress that these findings must be considered with nuance. 
The mere act of adding a monetization-related link or a cryptocurrency address in a video description certainly does not make a content creator more productive. 
However, adopting monetization strategies is likely to motivate higher productivity due to the creation of additional income streams (or at least the possibility of eventually creating them).
Another possibility is that, at a given point in time, creators decide to take their content production efforts more seriously, and when doing so, adopt alternative monetization strategies.
Although disentangling the exact causal structure between monetization and content production is hard, our results show that adopting alternative monetization strategies has a positive correlation with channel productivity. 
\section{Problematic channels' usage of alternative monetization strategies}

Adopting alternative monetization strategies is beneficial to a channel's productivity and growth, as it diversifies its revenue stream and decreases the likelihood of being impacted by  policy changes from YouTube.
Consequently, such strategies are of particular importance to channels producing problematic content.
As previously discussed, YouTube uses demonetization, i.e., losing ad revenue, as a punishment for content creators producing videos that violate YouTube's content policy~\cite{lunden_youtube_2018,youtube_borderline}.
In that context, creators that were punished or those who expect to be punished may leverage alternative monetization strategies to weaken the impact of demonetization.
Therefore, we conduct the following analyses using the \emph{Problematic Content} dataset, comprising 479 channels~(see Section~\ref{sec:data} for more details).
Our analyses are focused on three types of fringe content, Alt-light, Alt-right, and Manosphere.
We first compare the adoption patterns of alternative monetization strategies between the problematic and random channels
and report on domains that are disproportionally more popular among the problematic content creators.
Additionally, we analyze the amount of money problematic content creators earn from cryptocurrency donations, as well as from \domainformat{patreon}, the biggest player in the \textit{Donation} (\textbf{[DO]}) category.

\begin{figure}
    \includegraphics[width=\linewidth]{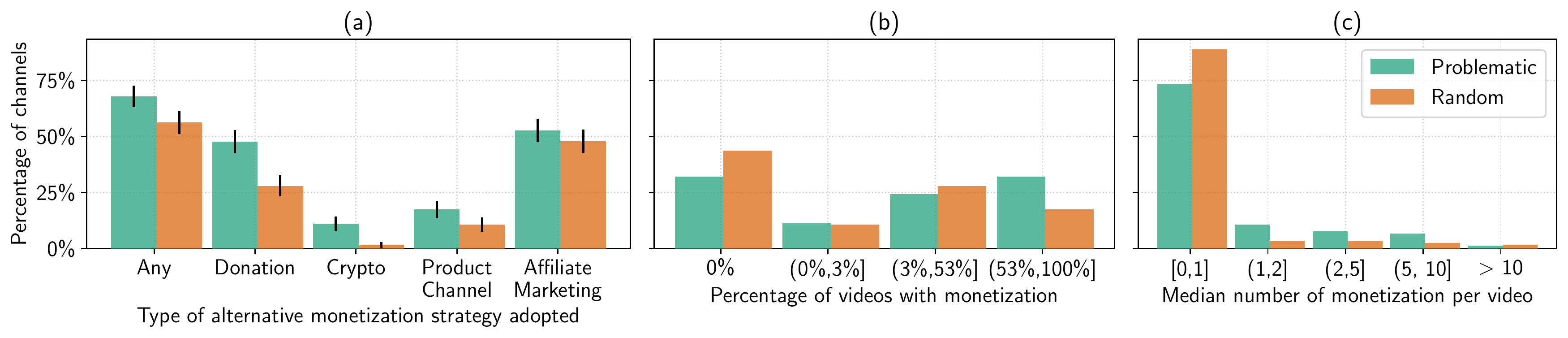}
    \caption{\label{fig:fringe} \textbf{Problematic channels and alternative monetization:} Comparison between problematic channels and random channels in terms of (a) the likelihood of adopting alternative monetization, (b) the percentage of videos with monetization links, groups are separated according to the quartile values of the monetization percentage. (c) the median number of monetization links in one video.
    In (b,c), the two distribution are significantly different~($p < 10^{-6}$, Mann--Whitney $U$-test). 
    Error bars in (a) represent 95\% CIs.}
\end{figure}

\xhdr{Alternative monetization practices of problematic channels}
We compare monetization practices between problematic channels and channels that we select randomly from the \emph{YouNiverse} dataset (henceforth referred to as random channels).
We follow a similar matching methodology as described in Section~\ref{sec:longevity}, matching one problematic channel with random channels according to their content, productivity, and the amount of attention received. For each year between 2017 to 2019, when problematic channels were the most active, we stratify channels into ten deciles according to the number of videos they produced and the number of views they received. Then, we consider two channels as matching candidates if and only if they started producing content from the same year,  are of similar channel category,
\footnote{
Channels in the \emph{Problematic Content} dataset don't have the content category information that was reported by the content creators~(see Section~\ref{sec:data}).
We thus examine the 196 channels in the \emph{Problematic Content} dataset that also show up in the \emph{YouNiverse} dataset and find that
60\% of them belong to one of the following three categories, News \& Politics, People \& Blogs and Entertainment.
Hence, to ensure that channels are of similar content, we only select channels that are from one of these three categories in the \emph{YouNiverse} dataset for the matching analysis (done with all the 479 problematic channels).
}
and, for every year from 2017 to 2019,  belong to the same stratum in terms of both number of videos produced and number of views received. Below, we report results from comparisons between how problematic and these carefully matched channels employ alternative monetization strategies, as illustrated in Figure~\ref{fig:fringe}(a).

First, we find that problematic channels are more likely to adopt alternative monetization.
Figure~\ref{fig:fringe}(a) compares the percentage of channels~($y$-axis) adopting different kinds of alternative monetization~($x$-axis). 
Around $68\%$ of the problematic channels adopt alternative monetization in at least one of their videos~(vs. $56\%$ for the counterparts).
In particular, they are far more likely to ask for a donation, either using services such as Patreon~($48\%$ vs. $28\%$), or using cryptocurrency~($11\%$ vs. $2\%$), and to sell products associated with the channel~($17\%$ vs. $11\%$).
There is little difference between the two types of channels when it comes to using affiliate marketing links.

Second, problematic channels also use alternative monetization strategies in more videos.
Figure~\ref{fig:fringe}(b) compares the percentage of videos with alternative monetization links (or cryptocurrency addresses) in their descriptions of problematic channels and their counterparts.
Again, the $x$-axis splits the percentage of videos including any alternative monetization strategy into four groups,
the group boundaries are determined using the quartile values of the distribution. 
The $y$-axis specifies the percentage of channels in each group.
Over $32\%$ of the problematic channels include alternative monetization links in more than $53\%$ of their videos (vs. $17\%$ for counterparts).
Also, only $32\%$ of problematic channels do not include alternative monetization links in any of their videos (vs. $44\%$ for counterparts).
The results remain the same when channels are split according to the quintile values or the decile values of the distribution.

Third, and lastly, problematic channels use a more diverse set of monetization strategies.
Figure~\ref{fig:fringe}(c) presents the median number of distinct links and cryptocurrency addresses adopted by a channel in their video descriptions.
We divide the median number of links and cryptocurrency addresses into five groups~($x$-axis), and report the percentage of problematic channels and their counterparts that fall under each group~($y$-axis).
Similar to what we observed in the previous plots, again, we find that problematic channels use monetization strategies more aggressively.
For instance, around $7\%$ of the problematic channels have between 6 to 10 different URLs with monetization domains or cryptocurrency addresses per video description in median~(vs. $2\%$ for the matched counterparts).

\xhdr{Donation domains for problematic channels}
Recently, many platforms, including Patreon, have taken action against producers of problematic content~\cite{patreon_qanon}.
In expectation of being taken down by mainstream platforms, problematic content creators might resort to a different set of alternative monetization services as compared to random channels.
For example, Figure~\ref{fig:fringe}(a) shows that problematic channels are far more likely to ask for cryptocurrency donations, potentially to avoid regulation.  
Alternative monetization through donation has been largely exploited by problematic channels.
Compared to alternative monetization through products sold by channels,
in which case each channel might have to host its own website to sell products,
the particular domains that are being used for donation are more likely to be platforms with multiple users,
hence having a larger impact.

\begin{table}
\small
\caption{\label{tab:fringe_domain}  Top ten donation domains with the largest percentage of channels that use them being problematic, computed on the combined dataset with both \emph{YouNiverse} and \emph{Problematic Content}. We mark in bold the domains with a higher probability of being used by problematic channels than the general distribution~(first row).}
\begin{tabular}{lr}
\toprule
Donation domain & \% of problematic channels\\
& using the domain\\
\midrule
All & 6.15\% \\
\midrule
feedthebadger.com&\textbf{85.7\%}\\
wesearchr.com     &\textbf{58.3\%}\\
hatreon.net       &\textbf{51.9\%}\\
makersupport.com  &\textbf{32.1\%}\\
subscribestar.com &\textbf{14.9\%}\\
projectveritas.com&\textbf{8.7\%}\\
venmo.com         &\textbf{6.7\%}\\
donorbox.org      &\textbf{6.6\%}\\
paypal.com        &1.4\%\\
gofundme.com      &1.0\%\\
\bottomrule
\end{tabular}
\end{table}

In Table~\ref{tab:fringe_domain}, we list the top ten donation domains that are more popular among problematic channels. 
To conduct this analysis, we first exclude all problematic channels from the \emph{YouNiverse} dataset.
Then, for each donation domain, we compute the ratio between channels using the specific domain being problematic and the total number of channels that have mentioned this domain at least once in one of its video descriptions.
Each domain has to be mentioned by at least 20 different channels~(0.01\% of all channels),
or at least 5 different problematic channels~(0.8\% of all problematic channels).
As a baseline, 6.15\% of the channels that ask for donations are fringe.
Most of the top domains that are above this baseline, \domainformat{feedthebadger.com}~\cite{honeybadger}, \domainformat{wesearchr.com}~\cite{cnbc}, \domainformat{hatreon.net}~\cite{hatreon}, \domainformat{makersupport.com}~\cite{makersupport}, and \domainformat{projectveritas.com}~\cite{projectveritas} are known to be used by creators of problematic content.
As of April, 2021, only \domainformat{feedthebadger.com} and \domainformat{projectveritas.com} are online. 
However, it is worth noting that some of the more mainstream services \domainformat{subscribestar.com}, \domainformat{venmo.com} and \domainformat{donorbox.org} are also above this baseline.

\xhdrNoPeriod{How much do problematic channels benefit from alternative monetization?}
For content creators, the more money they earn from alternative monetization strategies,
the less they have to rely on YouTube's shared ad revenue.
Hence, to understand how alternative monetization strategies may weaken the usage of demonetization as a moderation strategy, we explore how much problematic content creators have earned outside YouTube's partner program.
We focus on two specific alternative monetization strategies: gathering paid subscriptions on Patreon,  and asking for Bitcoin or Ethereum donations. 
As shown in Table~\ref{tab:major_domains}, \domainformat{patreon} is the most used domain in the \textit{Donation} category.
In $56.3\%$ of the cases when \textit{Donation} is used as a monetization strategy, a \domainformat{patreon} link was used.
Likewise, Bitcoin and Ethereum are the two of the three most frequently employed cryptocurrencies, accounting for $68\%$ of all cryptocurrency addresses found in video descriptions.
For this analysis, we collect additional data from Graphtreon,
\footnote{\url{https://graphtreon.com/}}
a website that tracks YouTube content creators' Patreon income over the years,
and from the blockchain history of Bitcoin and Ethereum.

\begin{figure}
    \centering
    \includegraphics[width=.7\linewidth]{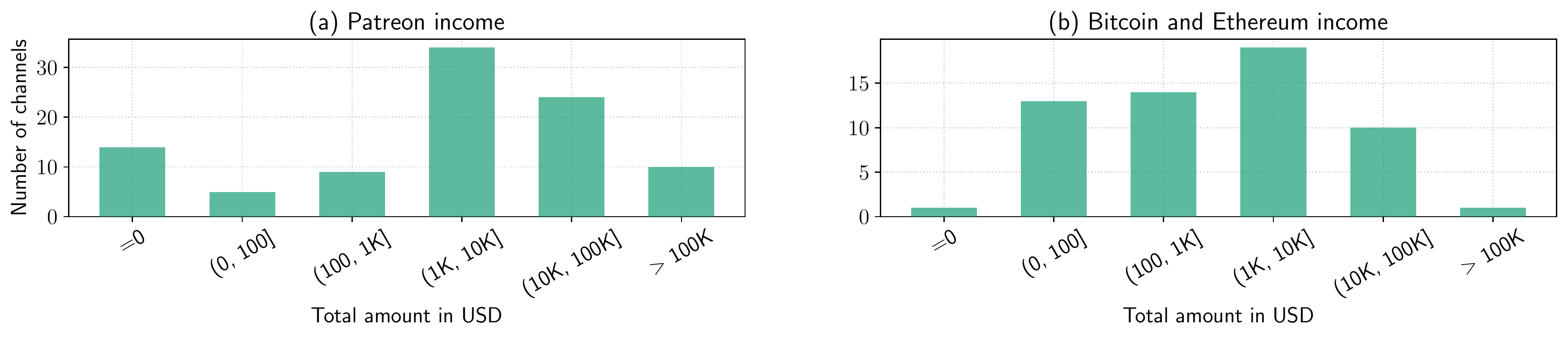}
    \caption{\label{fig:fringe_income} Problematic channel income (a) from Patreon~(97 channels), median income 5540 US dollars, (b) Bitcoin and Ethereum~(58 channels), median income 1155 US dollars.}
\end{figure}

We find that problematic channels benefit financially from alternative monetization in important ways.
Out of the $497$ problematic channels in the dataset,
$97$ mentioned at least one Patreon link in one of their video descriptions 
and $58$ mentioned at least one Ethereum or Bitcoin address in one of their video descriptions.
Figure~\ref{fig:fringe_income}(a) breaks down the lifetime income from all the Patreon links mentioned by problematic channels. 
Figure~\ref{fig:fringe_income}(b) breaks down the lifetime income until November 2020 from all the Bitcoin and Ethereum addresses mentioned by each channel. 
In both plots,
the $x$-axis specifies a total income level, ranging from earning nothing~(leftmost) to more than USD $100,000$ in total~(rightmost).
The $y$-axis specifies the number of channels with a given income level.
Most of the channels earned more than USD $1,000$ in total in both cases,
68 channels received more than USD $1,000$ on Patreon and 30 channels~(the three rightmost bars) received more than USD $1,000$ on Bitcoin and Ethereum addresses.

This analysis has some limitations.
It is possible that the Bitcoin and Ethereum addresses are used for other purposes other than receiving donations from YouTube audiences.
Patreon links in a channel's video might not be associated with the creators themselves,
but with someone else who wants to be promoted by this channel.
In both cases, we may overestimate the alternative monetization income for a channel.
Nevertheless, the analysis sheds light on how much problematic content creators may benefit from alternative monetization financially.
For example, Honey Badger Radio, one of the channels promoting Men's Rights Activists, is estimated to have earned over USD 260,000 on Patreon during its six-year presence on the platform.
A channel like this is likely to only receive minimal impact from a demonetization punishment, 
as they have another reliable source of income that is large enough to sustain themselves.

\section{Discussion}

Our results suggest that YouTube channels widely adopt a diverse set of alternative monetization strategies (\textbf{RQ1}). 
These strategies include affiliate marketing, the sale of products related to channels, and requests for donations through third-party platforms, or cryptocurrencies. 
In this context, we show that these strategies must be considered as an important monetary incentive and consideration when studying the platform. 
Our findings also indicate that there is a positive correlation between the adoption of alternative monetization and content production,
hinting that these strategies are indeed associated with changes in the content creation dynamics of the platform (\textbf{RQ2)}.
Lastly, our findings raise concerns about the ability of channels producing problematic content to exploit alternative monetization to avoid being punished due to policy violations (\textbf{RQ3)},
calling for more careful policy design. 
In the remainder of this section, we discuss the implications of our findings and possible future work.

\xhdr{Monetization and content creation}
Our work shows that alternative monetization is extremely prevalent among channels on YouTube and that it is likely to have a substantial impact on platform dynamics.
This is especially true for the creators, whose participation is essential to the platform~\cite{burgess2018youtube}.
Previous works studying the evolution of YouTube's ad-based revenue-share program~\cite{coromina2020follow,kopf2020rewarding,vonderau2016video} have pointed out that creators may optimize their content to receive more income~\cite{coromina2020follow} and that industry players may try to profit from the monetized content through multi-channel network~\cite{vonderau2016video}.
Alternative monetization strategies may change the YouTube ecosystem in a similar way.
For instance, creators may become more willing to produce content and change what kind of content they produce.

As suggested by previous work~\cite{munger2020right}, monetary incentives play an essential role in content production on YouTube. 
As such, the growing prevalence of alternative monetization practices is bound to impact the dynamics and motivations for content creation.
For example, alternatives such as Patreon, where creators can directly receive monetary contributions from their fans,
weaken the link between earnings and views.
Such contributions "incentivize the creation of a devoted fanbase and transform the revenue process into two-way communication between creator and audience"~\cite{munger2020right}.
Even if videos created by some creators receive only a few views, they can amass substantial earnings if a small but dedicated set of fans are willing to pay for a subscription.
This idea aligns with Kelly's 2008 essay ``1,000 True Fans,'' where, years before the advent of Kickstarter and Patreon, he toys with the idea that artists, producers, inventors and makers could make a living as long as they managed to find 1,000 true fans who  are willing to pay one day's wage per year to support them~\cite{kelly1000}.

Indeed, as pointed out by Gillespie, YouTube's business model and the affordances of the platform have consequences on what content creators make available~\cite{gillespie2010politics}, shaping the content that we see.
The forms of monetization allowed, "alternative" or not, can, therefore, be understood as part of the content moderation policy of the platform.
As creators shift to alternative monetization strategies, their content production goals may change. 
This interplay between content creation and monetization as moderation -- particularly for problematic content -- is an important future area of research that will build on existing work on platform dynamics on YouTube~\cite{lewis_this_2020, munger2020right, ribeiro_auditing_2020,cunningham2019creator}. 

\xhdr{YouTube as a gatekeeper} 
Our findings have significant implications for understanding YouTube's current role as a gatekeeper~\cite{gillespie2010politics, barzilai2008toward},
using on-platform monetization as a form of moderation.
For example, the platform prevents ads from being matched to videos containing problematic keywords and topics~\cite{markup}. 
In 2019, YouTube launched its new policy against
content that is at the borderline of violating the platform's policies~\cite{youtube_borderline}.
Videos under this category are not recommended to viewers and cannot display ads.
According to YouTube,
this policy (along with other improvements to their platform) has greatly reduced watch time on borderline content.
In a 2021 interview with the \textit{Washington Post,} the company's CTO claimed that only between 0.16\% and 0.18\% of views went to content that broke the platform rules, a 70\% reduction compared to 2017~\cite{wapost_youtube_2021}.
However, even if YouTube is effective at finding all the borderline content,
given the sheer size of the platform---one billion hours spent on the platform daily~\cite{youtube_hours}---0.16\% still amounts to millions of viewing hours.

While YouTube was likely successful in reducing exposure to problematic content, our findings highlight that platform-based demonetization might not be an effective moderation tool.
An extreme such example is the case of Alex Jones' YouTube channel, \emph{InfoWars}.
Before its ban in late 2018, the channel featured over 30,000 videos and gathered over 2 million subscribers~\cite{coaston_youtube_2018}.
Jones was not eligible to monetize his content during this period~\cite{lunden_youtube_2018}, but it did not matter:  the core of his revenue was centered around selling supplements and other products online, which amassed him millions of dollars each year~\cite{williamson_conspiracy_2018}.

Our results show that many of the content creators who focus on Alt-light, Alt-right, and Manosphere content, not only the high profile ones, benefit significantly from alternative monetization, which would allow them to keep producing without ad revenue.
Therefore, we argue that moderation through demonetization is not likely to be an effective tool in disincentivizing the production of problematic content, and may even result in a shift of content produced towards committed audiences, as noted above.
An important finding to consider here is that a few major players of alternative monetization strategies are extremely prevalent (cf.\ Fig.~\ref{tab:major_domains}).
Furthermore, some of the alternative monetization services are particularly popular among the problematic content creators that we study~(cf.\ Table~\ref{tab:fringe_domain}).
For example, around 15\% of the channels in our dataset using SubscribeStar produce content that we consider to be potentially problematic.
As our study on problematic channels was limited to three categories, it is likely that there exist more services that are heavily exploited by creators of policy-violating content. 
Nevertheless, these findings suggest that alternative monetization services should be evaluated as key platforms,
and be held to public scrutiny.
We note that some of these players do not have specific anti-hate or anti-harassment policies (e.g., Amazon's associate program~\cite{amazon_associate}),
and others have such policies but seem to enforce them inconsistently (e.g., Patreon~\cite{patreon2021community}).

While content creators producing problematic content adopt alternative monetization strategies more often than the carefully chosen comparison set of creators (cf. Fig.~\ref{fig:fringe}), our results show that all creators are increasingly adopting such strategies (cf. Fig.~\ref{fig:channel_adoption}).
These alternative monetization strategies 
form an important part of the growing creator economy~\cite{shapiro2018unlocking}.
Moreover, they allow creators to increase their financial gains, decrease their reliance on specific money streams (e.g., YouTube's partner program~\cite{christin_drama_2021}),
and, as noted above, challenge the platform's gatekeeping power~\cite{cunningham2019creator}.
These contributions need to be considered when arguing whether to disallow or to limit such strategies on YouTube or on other platforms.
Alternative monetization and the consequence it brings should be studied carefully to enable appropriate moderation.

\xhdr{Detecting alternative monetization}
Although our study mainly focuses on YouTube, 
many other platforms that support content distribution can be used for alternative monetization.
The affordances of YouTube, i.e., the video descriptions, make it especially easy for alternative monetization strategies to thrive,
as the links to alternative monetization services can be presented alongside the content.
However, this is not unique; other platforms like Facebook or the alternative video streaming platform Bitchute have similar affordances.
Further, users on Twitter, Instagram, or Gab can still mention alternative monetization links in their profiles,
encouraging their audiences to contribute.
In this paper, we developed a taxonomy of these strategies and a simple approach to detect them at scale.
We believe that this, along with the labeled set of monetization domains we release, may help researchers characterize the usage of such strategies in other platforms, as well as to better understand how they are employed in a way that impacts our information ecosystem.
These resources could also be helpful to social networking platforms, which ought to consider how websites like \textit{Patreon} and \textit{Amazon} help to shape the content that's fit to be created.

An interesting direction for future work is to study the interplay between "official" and "alternative" monetization strategies on YouTube itself. Researchers could build upon our work to study changes in the adoption of alternative monetization strategies when creators join YouTube's ad revenue sharing program or when they are banned from it. 
This analysis could help to further clarify YouTube's role as a gatekeeper.
In order to address these questions, we encourage YouTube to share data with trusted researchers on when or whether a channel or a piece of content is part of, or excluded from, monetization programs.
\section{Conclusion}

In this work,
we studied how content creators repurpose YouTube's affordances in order to monetize off the platform, using datasets that include both popular and controversial channels on YouTube.
We developed a taxonomy of alternative monetization using qualitative coding.
Further, we designed a bootstrapping algorithm to identify alternative monetization related domains and measured the usage of different alternative monetization strategies in our datasets. 
Our results show that alternative monetization is extremely prevalent among popular channels on the platform,
and has a positive correlation with content production.
Moreover, creators of problematic content, such as Alt-light, Alt-right, and Manosphere content, exploit alternative monetization strategies to a large extent.
We hope these findings may help future research better understand YouTube as a monetization platform and examine its role as a gatekeeper.
Lastly, with our developed taxonomy and methodology, platform providers including YouTube can investigate how alternative monetization might affect their policy making.
\section{Acknowledgements}
We thank Steven J. Jackson for his consultation on this work. 
This material is based upon work partially supported by the National Science Foundation under grants SaTC-2120651 and IIS-1840751. 
Manoel Horta Ribeiro is supported by a Facebook Fellowship Award.
Robert West is partly supported by a grant from the EPFL/UNIL Collaborative Research on Science and Society (CROSS) Program, the Swiss National Science Foundation (grant 200021\_185043), and the European Union (TAILOR, grant 952215), and gifts from Google, Facebook, and Microsoft.

\bibliographystyle{ACM-Reference-Format}
\bibliography{citation}
\newpage
\appendix

\begin{figure*}
\centering
\includegraphics[width=\linewidth]{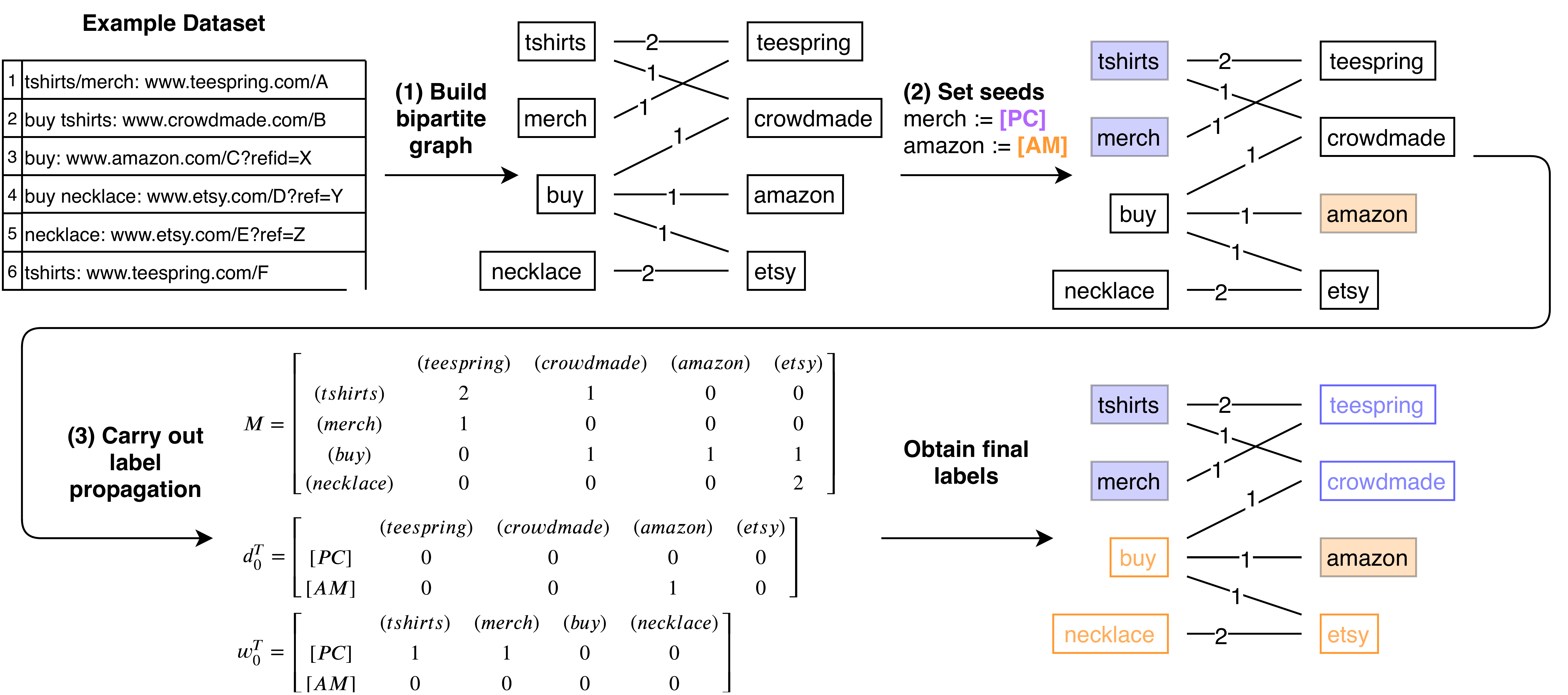}
\caption{\textbf{Summary of label propagation.}
We depict the different steps in our methodology for a toy dataset consisting of six (hypothetical) video descriptions. Note that each step \textbf{(1)}--\textbf{(3)} is explained in the main text.}
\label{fig:label_prop}
\end{figure*}

\section{Label Propagation Methodology and Validation}
\label{ap:method}

Our method consists of three steps, illustrated in Figure~\ref{fig:label_prop} and described below:

\xhdr{(1) Build bipartite graph} 
First, we build a word-domain co-occurrence bipartite graph $G(words,$ $domains)$ leveraging the datasets summarized in \tabref{tab:summary}. 
The $words$ nodes correspond to all the unigrams and bigrams that occur in at least $k$  different channels (we use $k$=30),
while the $domains$ nodes correspond to all the successfully-parsed domains from the millions of video descriptions in the datasets leveraged.
Once nodes are set, we create edges between a $word$ node and a $domain$ node if there is an instance in our dataset such that the word occurred before an URL belonging to the domain,
either in the same line or in the line right before the URL.
The weight of each edge captures the number of times this $\langle word,$ $domain\rangle$ pair co-occurred.

\xhdr{(2) Set seeds} 
Second, we define a set of ``seeds'' in this graph, assigning a small set of words and domains to one of the three categories of interest. Additionally, we also assign counter-examples: domains and words not related to monetization (which we refer to as seeds of the category \textbf{Other}). Seeds are obtained from the annotation process carried out when developing the taxonomy and are shown in Table~\ref{tab:seeds}.

\xhdr{(3) Label propagation} Lastly, we perform label propagation following the methodology inspired by the traditional label propagation algorithm \cite{zhu_introduction_2009}. 
Our method leverages the matrix $M$ of shape ($|words| \times |domains|$) derived from the bipartite graph obtained in step (1), and arrays $w$ and $d$, of shapes $|words| \times 4$ and  $|domains| \times 4$. 
For the arrays, each column corresponds to one of the categories (\textbf{[DO]}, \textbf{[PC]}, \textbf{[AM]}, and \textbf{Other}), and each row corresponds to either a word or a domain. 
We initialize rows corresponding to the seed words and domains: assigning the value $1$ to the row corresponding to its category, and $0$ for the remaining ones.
Then at each iteration $j$ we obtain new arrays $d_j$ and $w_j$ by:
\begin{enumerate}
    \item Multiplying the previous arrays  $d_{j-1}$ and $w_{j-1}$ by L1-normalized versions of $M$. We refer to it as $M_{domain}$, where each column sums to 1, and  $M_{word}$, where each row sums to 1: 
    \begin{align*}
    d_j = M_{domain}^T  \cdot w_{j-1} \quad \text{and} \quad
    w_j = M_{word} \cdot d_{j-1} 
    \end{align*}

    \item L1-normalizing the arrays $d_j$ and $w_j$ so that each column sums to 1.
    
    \item ``Clamping'' the arrays $d_j$ and $w_j$: for rows corresponding to the seed words and domains: we assign the value $1$ to the row corresponding to its category, and $0$ for the remaining ones.
\end{enumerate}
Convergence of this method is guaranteed~\cite{zhu_introduction_2009}.
We repeat this process until we have a small change between arrays $d_j$ and $d_{j-1}$ ($|d_j - d_{j-1}| < 10^{-15}$).
Final arrays $d$ and $w$ contain the inferred labels for all words and domains in the graph. We obtained 736 domains related to donations (\textbf{[DO]}), 36,958 domains related to affiliate marketing (\textbf{[AM]}), 1565 domains related to products sold by the channel (\textbf{[PC]}) and 136,161 domains not related to monetization (\textbf{[NM]}).

\begin{table*}[t]
{
\caption{\textbf{Seed words and domains.} For each  category, we depict the seed words and domains used. }
\label{tab:seeds}
{
\centering

\begin{tabular}{p{1.2cm}|p{2cm}|p{3cm}|p{2.25cm}|p{3.5cm}}
\toprule
 & \textbf{[DO]} & \textbf{[PC]}  & \textbf{[PO]} & \textbf{[NM]}    \\ \midrule
\small{Word seeds} & \wordformattable{donate}, \wordformattable{donation}, \wordformattable{stream labs}, \wordformattable{paypal} & \wordformattable{merch}, \wordformattable{merchandise}, \wordformattable{shirt}, \wordformattable{shirts}, \wordformattable{swag}, \wordformattable{merch shop}, \wordformattable{mugs}, \wordformattable{mug}, \wordformattable{coaching}, \wordformattable{course}, \wordformattable{courses}, \wordformattable{lesson}, \wordformattable{lessons} & \wordformattable{promo code}, \wordformattable{code}, \wordformattable{discount} & \wordformattable{intro}, \wordformattable{follow instagram}, \wordformattable{facebook}, \wordformattable{soundcloud}, \wordformattable{snapchat}, \wordformattable{twitch} \\\midrule
\small{Domain seeds} & \domainformattable{paypal}, \domainformattable{subscribestar}, \domainformattable{patreon}, \domainformattable{gofundme}, \domainformattable{ko-fi} & \domainformattable{teespring}, \domainformattable{bonfire}, \domainformattable{merchlabs}, \domainformattable{represent}, \domainformattable{crowdmade}  & \domainformattable{amazon}, \domainformattable{etsy}, \domainformattable{ebay}, \domainformattable{skillshare}, \domainformattable{squarespace}, \domainformattable{thegreatcourses} & \domainformattable{facebook}, \domainformattable{deviantart}, \domainformattable{twitter}, \domainformattable{cnn}, \domainformattable{dropbox}, \domainformattable{washingtontimes}, \domainformattable{pastebin}, \domainformattable{videvo}, \domainformattable{audiomack}, \domainformattable{canva}, \domainformattable{t}, \domainformattable{soundcloud}, \domainformattable{tiktok}, \domainformattable{freesfx}, \domainformattable{tiktok}, \domainformattable{imgur}, \domainformattable{pinterest}, \domainformattable{wired}, \domainformattable{snapchat}, \domainformattable{pnas}, \domainformattable{tapas}, \domainformattable{washingtonpost} \\ \bottomrule
\end{tabular}
}
}

\end{table*}

\xhdr{Validation}
We validated our results by manually annotating samples. We performed the task both at the 
domain level as well as at the URL level:

\begin{itemize}
\item \textbf{Domain-level validation}
For our first validation effort, we manually annotated three samples containing 500 domains in total. 
Two authors of this paper individually inspected each domain, analyzing the top 5 words most frequently used with it, and judged whether the label given was correct or false.
To obtain final labels, the authors then discussed disagreements individually.
The first sample (\texttt{rstrat}) contained, for each possible class, 50 random domains labeled by the method as the given class (totaling 200 domains).
The second sample (\texttt{pstrat}) contained the 50 most popular domains labeled by our method as each given class (again, totaling 200 domains).
Lastly, the third sample (\texttt{random}) contained 100 randomly sampled domains.
The rationale behind these different sampling mechanisms is to analyze how our methodology performed in different scenarios.
Results are reported in Table~\ref{tab:val}a-b.
Additionally, Table~\ref{tab:tab_val} reports random sub-samples ($n$=10 for each category/sample).

\item \textbf{URL-level validation}
For our second validation effort, we manually annotated a sample of 400 videos.
To obtain this sample, for each URL-related category,\footnote{Product Channel, Affiliate Marketing, Donation and Other; note that Cryptocurrencies were not validated this way because matching is exact.} we sampled 100 random videos such that at least one URL in the video description linked to the domain that is labeled with the category. 
Then, two authors of this paper individually inspected each video description and annotated whether the first URL in the video description that points to the domain in question was aligned with the label.
To obtain final labels, the authors then discussed disagreements individually.
Note that this criterion is stricter than what was used for the domain-level validation. For example, amazon.com is a domain commonly associated with Affiliate Marketing, and was annotated as such in the domain-level validation. Yet, sometimes users can share Amazon links that are not Affiliate Marketing,
for example, to advertise for a book that they have written.
The domain-level annotation does not capture this nuance, while the URL-level annotation does.
Results are reported in Table~\ref{tab:val}c.

\end{itemize}

\begin{table}[]
\centering
\footnotesize
\caption{\textbf{Validation Results.} We depict the result of our validation efforts at the domain level (in \textit{a} and \textit{b}) and at the URL level (in \textit{c}).
 }

\label{tab:val}
\begin{subtable}{0.7\linewidth}
\centering
\caption{Precision of our method on different random samples according to the domain-level validation.}
\begin{tabular}{r|c |llll}
\toprule
Sample & $n$ & \textbf{[DO]} &\textbf{[PC]} & \textbf{[AM]} & \textbf{[NM]} \\ \midrule
\texttt{rstrat} & 200 & 26\% & 100\% & 92\% & 68\% \\
\texttt{pstrat} & 200 & 60\% & 88\% & 100\% & 92\% \\
\texttt{random} & 100 & \multicolumn{4}{c}{\dotfill 84\%\dotfill} \\ \bottomrule
\end{tabular}
\vspace{5mm}
\end{subtable}%
\hfill
\begin{subtable}{0.475\linewidth}
\centering
\caption{Confusion matrix for the \texttt{pstrat} sample showing when and how our method makes mistakes in the domain-level validation.}
\begin{tabular}{r|llll}
\toprule
\diagbox{Real}{Pred.}  &  \textbf{[DO]} &\textbf{[PC]} & \textbf{[AM]} & \textbf{[NM]} \\
\midrule
\textbf{[DO]}      &        30 &              0 &             0 &       3 \\
\textbf{[PC]} &         0 &             44 &       0       &       0 \\
\textbf{[AM]}  &         0 &              4 &            50 &       3 \\
\textbf{[NM]}        &        20 &              2 &             0 &      44 \\
\bottomrule
\end{tabular}

\end{subtable}%
\hfill
\begin{subtable}{0.475\linewidth}
\centering
\caption{Confusion matrix for the URL-level validation showing when and how our method makes mistakes.}

\begin{tabular}{r|llll}
\toprule
\diagbox{Real}{Pred.}  &  \textbf{[DO]} &\textbf{[PC]} & \textbf{[AM]} & \textbf{[NM]} \\ \midrule
\textbf{[DO]} & 98 & 0 & 0 & 0 \\
\textbf{[PC]} & 1 & 99 & 18 & 5 \\
\textbf{[AM]} & 0 & 0 & 75 & 1 \\
\textbf{[NM]} & 1 & 1 & 7 & 94 \\ \bottomrule

\end{tabular}
 \end{subtable}

\end{table}

\xhdr{Shortcommings and post-processing}
As shown in Table~\ref{tab:val}a-b, our method performs particularly poorly  (\textbf{[DO]}) for donation domains in both the \texttt{pstrat} and the \texttt{rstrat} samples.
Our manual analysis reveals that our method struggles to distinguish between domains related to donations to streamers (e.g., \domainformat{patreon.com}) and those that are related to charity (e.g., \domainformat{orangutan.com}).
In the \texttt{pstrat} sample, for instance, out of the 20 times the method wrongly assigned \textbf{[DO]} to a domain in the \texttt{pstrat} sample, in 15 cases, the domain belongs to some charity.
We speculate that this happens because the set of words used to describe these two different kinds of domains is indistinguishable at times (\wordformat{donate}, \wordformat{support}, etc).
Fortunately, however, donation-related domains (\textbf{[DO]}) are the category with the fewest assigned domains (731). Moreover, most of the domains assigned in this category occur very sparsely, for example, 62\% of them occur in only 1 or 2 distinct channels. 
Note that the top 50 most popular donation domains cover over 95\% of all usages of donation domains. 
Thus, to address this issue, we manually correct the labels of the top 50 most popular domains labeled as donations,
improving the quality of annotation altogether.

After correcting for this issue, the results shown in  Table~\ref{tab:val}c, show that our model is highly accurate at the URL level.
In total, our method achieves $91.5\%$ accuracy.
However, we also find that our methodology may fail to distinguish between the \textit{Affiliate Marketing} category (\textbf{[AM]}) and the \textit{Products Channel} category (\textbf{[PC]}), predicting the later as the former.
False positive cases typically occur when the channel creator promotes a product by the channel using common shopping websites,
for example, selling a book that they have written on Amazon.
We find no good solution for this issue.
This bias might lead to an over-estimation of the \textit{Affiliate Marketing} category (\textbf{[AM]}) and a under-estimation of the \textit{Products Channel} category (\textbf{[PC]}) in our analyses.

\begin{table*}[]

{
\caption{\textbf{Showcase of results.} For each stratum (shown in the rows) and each category, (shown in the columns), we depict 10 randomly sampled domains obtained through our method. Mistakes are \underline{underlined}. For the \texttt{random} sample, we instead show 40 randomly sampled domains (out of the 100 domains in the sample).}
\label{tab:tab_val}
\small
\begin{tabular}{p{1cm}|p{3cm}|p{2.5cm}|p{3cm}|p{2.5cm}}
\toprule
 & \textbf{[DO]} & \textbf{[PC]}  & \textbf{[PO]} & \textbf{[NM]}    \\ \midrule
\texttt{rstrat} &  \domainformattable{vrdonate}, \domainformattable{\underline{muslimaid}}, \domainformattable{\underline{theintrepidfoundation}}, \domainformattable{\underline{kilmanjaromusic}}, \domainformattable{\underline{whatis36}}, \domainformattable{\underline{menageriecoffee}}, \domainformattable{freakinrad}, \domainformattable{\underline{sportable}}, \domainformattable{beyondtype1}, \domainformattable{\underline{jedfoundation}} &  \domainformattable{campuscustoms}, \domainformattable{everythingcerti}, \domainformattable{danielhowell}, \domainformattable{youinkit}, \domainformattable{jumpstartaffiliate}, \domainformattable{contextualelectronics}, \domainformattable{bloggerworkshop}, \domainformattable{wemakeyoulaughfilms}, \domainformattable{allinmerch}, \domainformattable{starsnipemerch} &  \domainformattable{supernutritionacademy}, \domainformattable{mezeaudio}, \domainformattable{fishinglyn}, \domainformattable{pulsateathleticwear}, \domainformattable{jerseysfc}, \domainformattable{babybykyra}, \domainformattable{scarletimprint}, \domainformattable{glamorhairlondon}, \domainformattable{elitemetaltools}, \domainformattable{disneysprings} &                                     \domainformattable{toskanaferien}, \domainformattable{modjunkiez}, \domainformattable{tes}, \domainformattable{\underline{thedrunkentaoist}}, \domainformattable{\underline{lisajblog}}, \domainformattable{yungeldr}, \domainformattable{frasiersterlingjewelry}, \domainformattable{cascaderecords}, \domainformattable{\underline{sennheiser}}, \domainformattable{\underline{izaha}} \\ \midrule
\texttt{pstrat} &                                                                                  \domainformattable{mycause}, \domainformattable{hatreon}, \domainformattable{\underline{onelist}}, \domainformattable{youcaring}, \domainformattable{duapps}, \domainformattable{\underline{unitedwayhouston}}, \domainformattable{pitchinbox}, \domainformattable{\underline{orangutan}}, \domainformattable{flattr}, \domainformattable{matcherino} &                          
\domainformattable{teepublic},\domainformattable{endoapparel}, \domainformattable{\underline{wpengine}}, \domainformattable{danandphilshop}, \domainformattable{eivor}, \domainformattable{myteespring}, \domainformattable{mrfijiwiji}, \domainformattable{postmaloneshop}, \domainformattable{shopredhare}, \domainformattable{fanfiber} &                                                   \domainformattable{elfcosmetics}, \domainformattable{urbanoutfitters}, \domainformattable{romwe}, \domainformattable{forever21}, \domainformattable{bestbuy}, \domainformattable{magik}, \domainformattable{colourpop}, \domainformattable{nordstrom}, \domainformattable{sephora}, \domainformattable{homedepot} & \domainformattable{steampowered}, \domainformattable{bandcamp}, \domainformattable{github}, \domainformattable{lnk}, \domainformattable{mega}, \domainformattable{gleam}, \domainformattable{app}, \domainformattable{myspace}, \domainformattable{strawpoll}, \domainformattable{spotify} \\ \midrule
\texttt{random} & \multicolumn{4}{p{12cm}}{
\domainformattable{bloggerworkshop}, \domainformattable{tedyandreas}, \domainformattable{senreve}, \domainformattable{tactics}, \domainformattable{sfbags}, \domainformattable{primerpeak}, \domainformattable{glamorhairlondon}, \domainformattable{lorac}, \domainformattable{orbxdirect}, \domainformattable{hisonjetski}, \domainformattable{jerseysfc}   \domainformattable{thejakartapost}, \domainformattable{\underline{sennheiser}}, \domainformattable{njstreetworkout}, \domainformattable{pocoyo}, \domainformattable{yesterdayorigins}, \domainformattable{sixblindkids}, \domainformattable{viridianmusic}, \domainformattable{yungeldr}, \domainformattable{onetiredworkingmommy}, \domainformattable{cascaderecords}, \domainformattable{cqcbmachinery}, \domainformattable{drnikkistarr}, \domainformattable{xboxachievements}, \domainformattable{nextschool}, \domainformattable{toptiertactics}, \domainformattable{aqicn}, \domainformattable{bootdiskrevolution}, \domainformattable{\underline{businessesforsale}}, \domainformattable{shawngmusic}, \domainformattable{\underline{evilcontrollers}}, \domainformattable{jwmusic}, \domainformattable{stormfreerun}, \domainformattable{\underline{puregold}}, \domainformattable{khoasellsflorida}, \domainformattable{\underline{izaha}}, \domainformattable{\underline{partyideasuk}}, \domainformattable{hivesforhumanity}, \domainformattable{\underline{yurptm}}, \domainformattable{wt1}} \\
\bottomrule
\end{tabular}
}

\end{table*}

\clearpage
\newpage

\begin{figure}[t]
    \centering
    \includegraphics[width=0.925\linewidth]{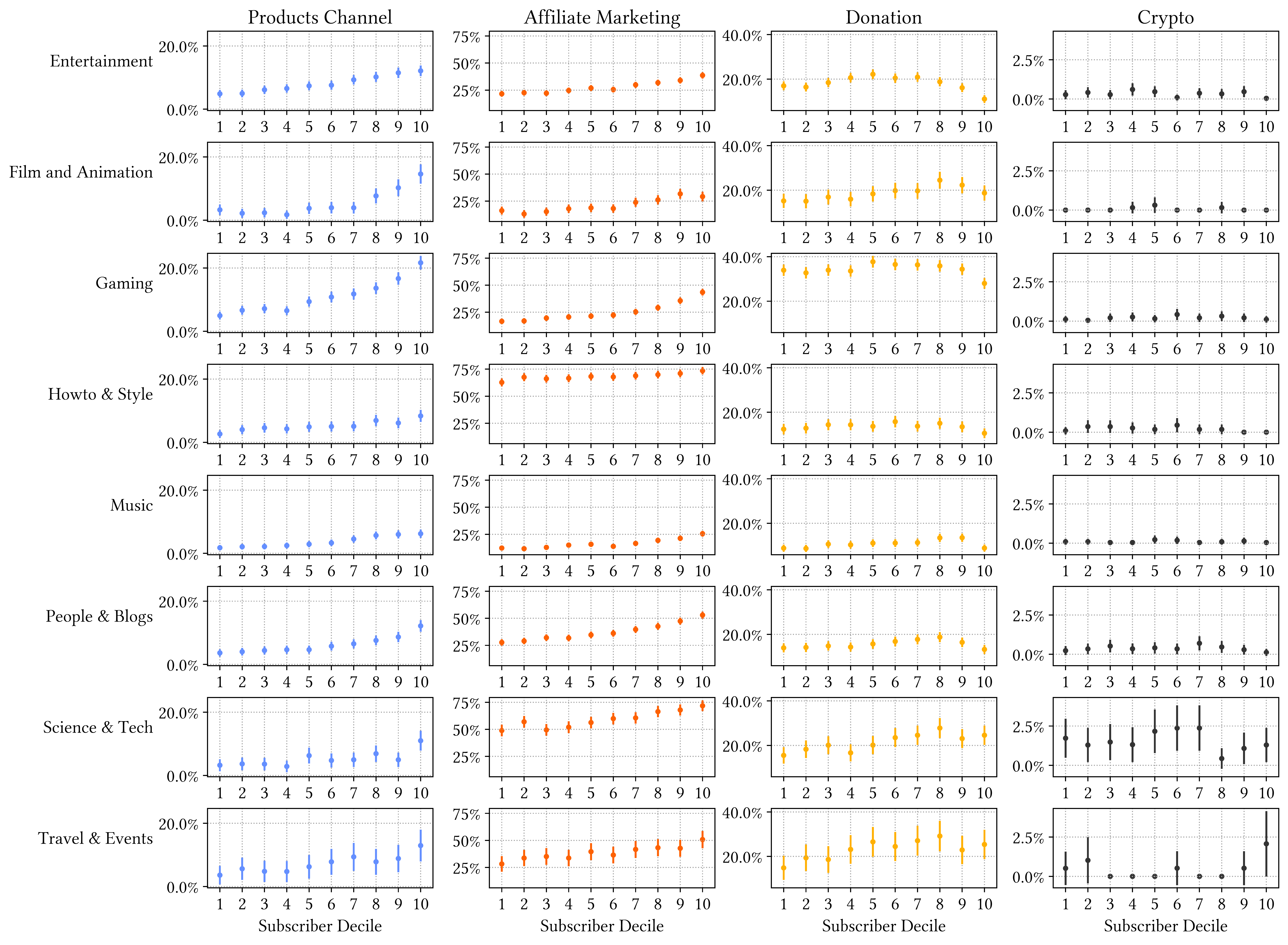}
    \caption{\textbf{Popularity and alternative monetization for different categories}: we repeat the analysis done in Figure~\ref{fig:subs} considering different channel categories separately.}
    \label{fig:subs_cats}
\end{figure}

\begin{figure}[t]
    \centering
    \includegraphics[width=0.925\linewidth]{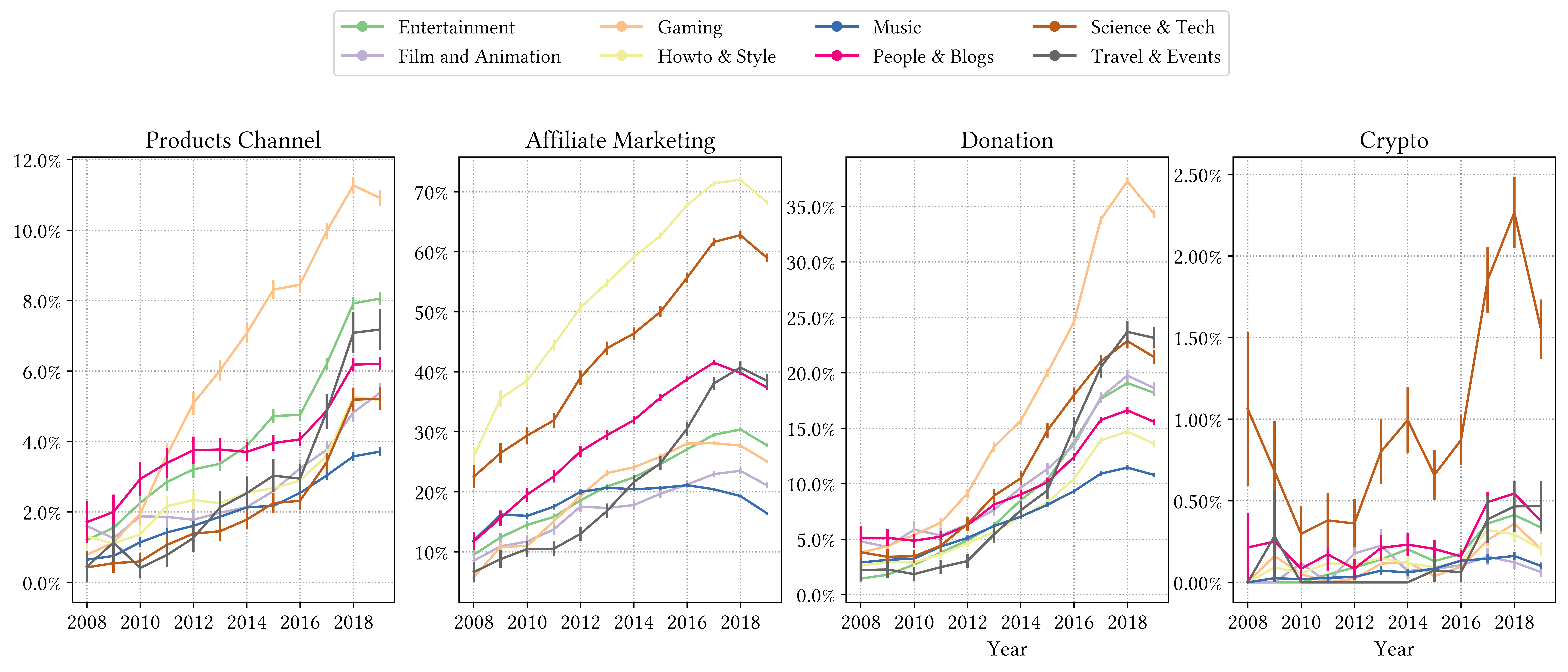}
    \caption{\textbf{Prevalence of alternative monetization through the years for different categories}: we repeat the analysis done in Figure~\ref{fig:channel_adoption}a considering different channel categories separately.}
    \label{fig:prev_cats}
\end{figure}
\end{document}